\documentclass[11pt]{article}

\usepackage{acl}

\usepackage{times}
\usepackage{latexsym}
\usepackage{subcaption}
\usepackage[ruled,vlined,linesnumbered]{algorithm2e}
\usepackage{amsmath}
\usepackage{booktabs}   
\usepackage{siunitx}    
\usepackage{multirow}
\usepackage{titletoc}
\usepackage[table]{xcolor}
\usepackage{makecell}   
\usepackage{tabularx}   
\usepackage{amssymb}
\usepackage{amsthm}

\newtheorem{theorem}{Theorem}[section]
\newtheorem{proposition}[theorem]{Proposition}

\usepackage[T1]{fontenc}
\usepackage{xcolor}
\definecolor{chatPurple}{RGB}{150, 130, 160}

\usepackage[most]{tcolorbox}

\newtcolorbox{ChatBox}[2][]{
    enhanced,                
    colback=white,           
    colframe=chatPurple,     
    coltitle=white,          
    fonttitle=\bfseries,     
    title={#2},              
    arc=3mm,                 
    drop shadow,             
    boxrule=0.5mm,           
    fontupper=\ttfamily\small, 
    left=3mm, right=3mm, top=3mm, bottom=3mm, 
    #1                       
}

\usepackage{graphicx}

\usepackage[utf8]{inputenc}

\usepackage{microtype}

\usepackage{inconsolata}

\usepackage{graphicx}

\title{MACRO-LLM: LLM-Empowered Multi-Agent Collaborative Reasoning under Spatiotemporal Partial Observability}

\author{
 \textbf{Handi~Chen},
 \textbf{Running~Zhao\textsuperscript{*}},
 \textbf{Xiuzhe~Wu\textsuperscript{*}},
  \textbf{Zhanfeng~Xu},
 \textbf{Edith~C.H.~Ngai\textsuperscript{\dag}}
 \\
 The University of Hong Kong,
\\
 \small{\{hdchen, rnzhao, xzwu, zhanfeng.xu\}@connect.hku.hk, chngai@eee.hku.hk
 }
}

\begin{document}
\maketitle

\begingroup
  \renewcommand\thefootnote{}
  \footnotetext{\textsuperscript{*}Equal contribution; \textsuperscript{\dag}Corresponding author.}
\endgroup

\begin{abstract}
Large Language Model (LLM) agents in distributed decision-making scenarios often operate as spatially separated entities and coordinate from fragmented local information. However, this physical dispersion constrains agents to limited local perception and finite temporal horizons. We characterize this bottleneck as spatiotemporal partial observability, in which spatial and temporal limitations are fundamentally coupled. Resolving spatial conflicts requires temporal reasoning about neighbors' future actions, while temporal planning requires spatial context beyond local perception. To bridge this gap, we introduce \textbf{MACRO-LLM}, \underline{LLM}-empowered \underline{m}ulti-\underline{a}gent \underline{c}ollaborative \underline{r}easoning under spatiotemporal partial \underline{o}bservability. The architecture integrates spatial and temporal reasoning within each decision cycle via three interdependent modules: (1) the CoProposer mitigates \textit{temporal uncertainty} by screening candidate actions via LLM-predicted rollouts; (2) the Negotiator overcomes \textit{spatial myopia} by resolving conflicts through mean-field statistical aggregation, grounded in the CoProposer's LLM-estimated rollout rewards; and (3) the Introspector closes the reasoning loop by analyzing environmental drift and attributing performance changes to refine strategies. Extensive evaluations on two complex long-horizon tasks, cooperative platoon planning and pandemic control, demonstrate that our framework enables robust coordination under spatiotemporal partial observability.
\end{abstract}

\section{Introduction}

\begin{figure}[!t]
    \centering
    \includegraphics[width=0.97\linewidth]{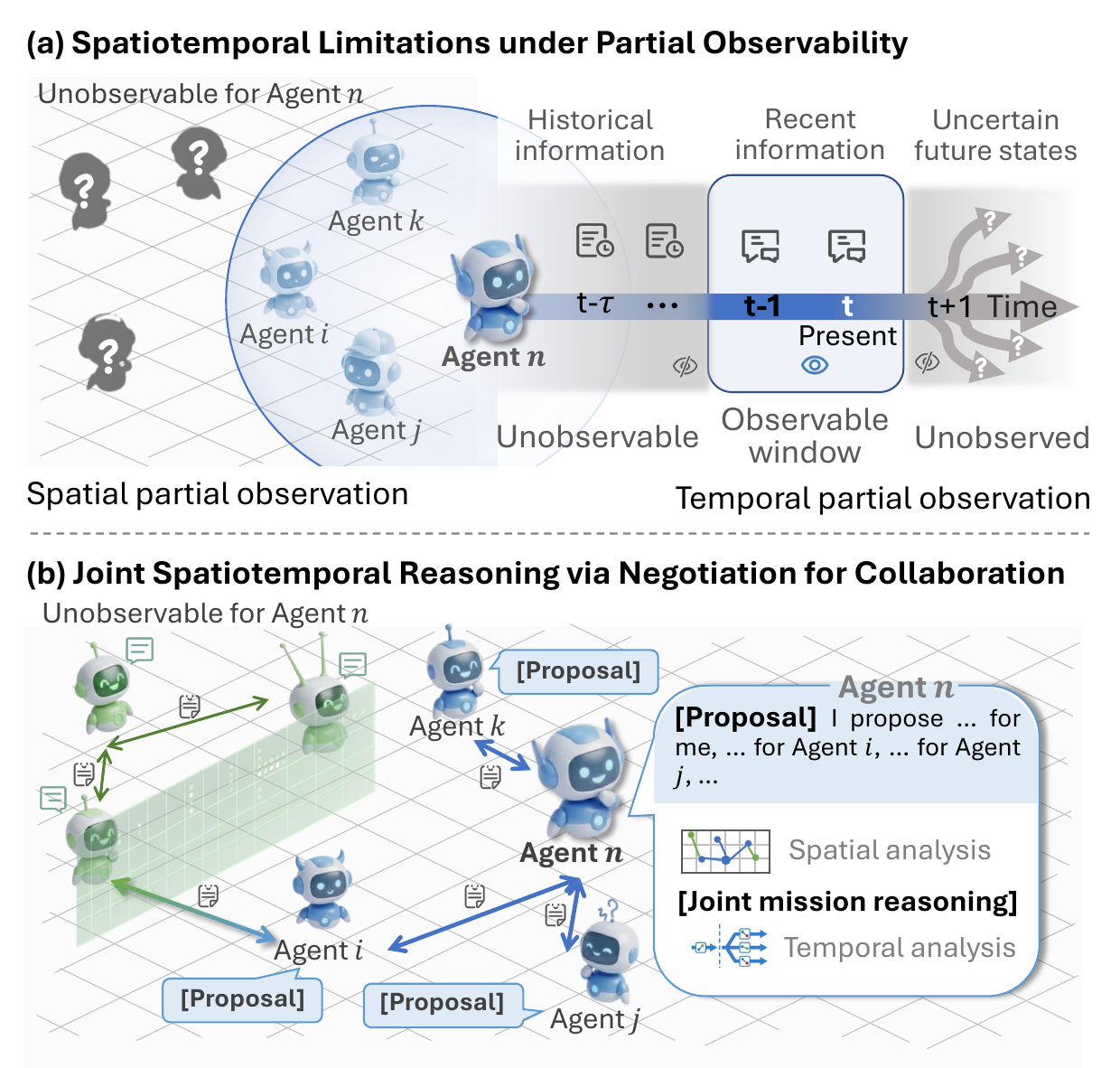}
    \caption{Illustration of spatiotemporal partial observability and the proposed collaborative reasoning framework. (a) Agents are constrained by limited local perception (spatial) and future uncertainty coupled with finite context windows (temporal). (b) Agents mitigate these limitations by exchanging proposals to establish a collaborative framework via neighborhood negotiation.}
    \label{fig:teaser}
\end{figure}

Large Language Model (LLM) agents have demonstrated remarkable human-like capabilities in problem-solving~\cite{problemsolving,cui2025curie}. While these agents successfully automate tasks across domains such as healthcare~\cite{personalhealth} and education~\cite{noteit, shi2025educationq}. However, individual agents often struggle with complexity and scalability in long-horizon scenarios. Inspired by the team collaboration in human societies, researchers are actively exploring systems where multiple LLM agents coordinate to achieve collective goals~\cite{Li2023CAMEL,agentcoll}. This collective intelligence amplifies individual capabilities, enabling the effective resolution of complex reasoning tasks through coordination and negotiation~\cite{surveyagent}. However, most existing frameworks assume shared global context or unrestricted communication among agents. 
As LLM agents are increasingly considered for distributed decision-making, coordination under limited local perception remains a key challenge.

In such settings, agents operate as distributed entities~\cite{chen2025toward}, which inherently impose spatiotemporal partial observability.
Physical dispersion constrains agents to limited local perception, while the LLM's fixed context window restricts their historical temporal information. This challenge arises in many scenarios. In pandemic control, districts must coordinate containment policies with only local infection data. 
In strategic traffic coordination, agents reason about flow-level decisions under communication limits and privacy constraints. Addressing partial observability in multi-agent reasoning is important to improve the efficiency and effectiveness of collaboration.

Specifically, partial observability manifests along two dimensions: \textit{spatial} and \textit{temporal}, as illustrated in Fig. \ref{fig:teaser}(a). \textit{Spatially}, constrained by communication limits and environmental complexity, each agent perceives only a fraction of the global information, preventing it from inferring the full context of the collective task. \textit{Temporally}, the stochastic environment and unpredictable behaviors of others make long-term consequences difficult to predict, while finite historical context further hinders alignment of short-term decisions with long-term objectives. Together, these constraints result in suboptimal or even conflicting actions that degrade collective efficiency and effectiveness.

Existing LLM-based Multi-Agent Systems (MAS) aim to expand their spatial perception through inter-agent communication, as shown in Fig. \ref{fig:teaser} (b). This is typically implemented via hierarchical structures that aggregate information to a central node~\cite{yu2024fincon, estornell2025train} or fully-connected graphs that construct a pseudo-global view~\cite{chan2023chateval}. However, these methods often incur communication bottlenecks at central nodes, limiting their scalability in large-scale, dynamically evolving environments. Temporally, while memory modules help mitigate limited historical observations~\cite{generativeagents,zhang2024building}, the inherent unpredictability of future states remains unaddressed. This uncertainty makes it difficult for agents to align immediate actions with long-term goals and capture cascading temporal effects.
Prior to LLM-based MAS, Multi-Agent Reinforcement Learning (MARL) provides a traditional framework for partial observability but suffers from poor generalization and excessive training costs~\cite{huh2023multi}, limiting its feasibility 
for complex real-world applications.


In this paper, we propose \textbf{MACRO-LLM}, an \underline{LLM}-empowered \underline{m}ulti-agent \underline{c}ollaborative \underline{r}easoning under spatiotemporal partial \underline{o}bservability, evaluated across two distinct simulator domains. A key insight is that spatial and temporal partial observability are \emph{coupled}: resolving spatial conflicts requires predicting how neighbors will act in the future, while planning over time requires spatial context beyond local perception. Existing approaches address these dimensions in isolation. MACRO-LLM bridges this gap with a negotiation-centric architecture that integrates spatial and temporal reasoning within each decision cycle. Three interdependent modules support this integration. The CoProposer generates collaborative proposals and evaluates them using LLM-predicted rollouts to mitigate temporal uncertainty. The Negotiator uses mean-field statistical aggregation over unobservable agents to resolve conflicts and overcome spatial myopia. The Introspector refines strategies through scenario-aware adaptive reflection, enabling continuous adaptation. Our main contributions are summarized as follows:

\noindent\textbf{Insightful.}
We explicitly decompose partial observability into spatial and temporal dimensions and address them jointly through a unified decentralized negotiation framework that couples semantic reasoning with mean-field statistical features, enabling robust long-horizon coordination without central coordinators or global views.

\noindent\textbf{Technical.}
We develop MACRO-LLM with three interdependent modules forming a closed reasoning loop: the Negotiator's confidence assessment consumes the CoProposer's LLM-estimated rollout rewards, and the Introspector's revision signals reshape the strategies for subsequent cycles. This interdependence constitutes the core contribution beyond the individual techniques each module builds upon.

\noindent\textbf{Experimental.}
We evaluate on two complementary domains, Cooperative Platoon Planning (CPP) and Pandemic Coordination (PC), covering continuous and discrete actions, linear and graph topologies, fine-grained and long-horizon planning. Results demonstrate MACRO-LLM's superior coordination performance, scalability, and robustness. 
\section{Related Works}

\paragraph{LLM-based Multi-Agent Systems.}
Existing LLM-based MAS have evolved from general collaboration frameworks~\cite{Li2023CAMEL, Wu2023AutoGen} to structured systems with persistent memory and reasoning capabilities~\cite{Hong2023MetaGPT, Wang2023Voyager}. Beyond general-purpose collaboration, LLM-based negotiation has also been instantiated in specific domains, such as autonomous driving, where agents negotiate over perception and trajectory-level decisions \cite{liu2025colmdriver}.
To address spatial partial observability, recent works adopt either \textit{hierarchical aggregation}, which centralizes information flow~\cite{yu2024fincon, estornell2025train}, or \textit{belief-based inference}, which deduces hidden states via reasoning chains~\cite{li2023theory, davidson2024evaluating}. 
However, these approaches often rely on ideal communication bandwidth or static topologies, failing to account for the dynamic coupling of spatial dispersion and temporal uncertainty. 
Additionally, swarm-based optimization methods~\cite{feng2024model, zhang2025swarmagentic} evolve agent configurations through offline iterative search, yielding static configurations that cannot adapt to runtime dynamics or resolve conflicts under partial observability.
MACRO-LLM targets decentralized reasoning under spatiotemporal partial observability and evaluates a shared coordination architecture across two distinct tasks.

\paragraph{Multi-Agent Reinforcement Learning.} 
Traditional approaches address partial observability via the decentralized partially observable Markov decision process framework~\cite{oliehoek2016concise, koops2024approximate}, often utilizing centralized training with decentralized execution \cite{kia2024memory, bernstein2005bounded} or mean-field approximations \cite{yang2018mean, ganapathi2020multi} to enhance scalability. While algorithms like MAPPO \cite{schulman2017proximal, wu2021coordinated}, DMPO \cite{ma2024efficient}, and communication-aware methods like IC3Net \cite{ic3net2018learning} improve stability, they remain computationally intensive and environment-specific. They require costly retraining when transferring to new scenarios due to poor generalization~\cite{huh2023multi, oroojlooy2023review}. MACRO-LLM reuses the same coordination architecture across distinct tasks without LLM parameter updates.

Extended discussions of these research directions are provided in \textbf{Appendix}~\ref{app:related_works}.
\section{MACRO-LLM Framework}
\begin{figure*}[!t]
    \centering
    \includegraphics[width=1\linewidth]{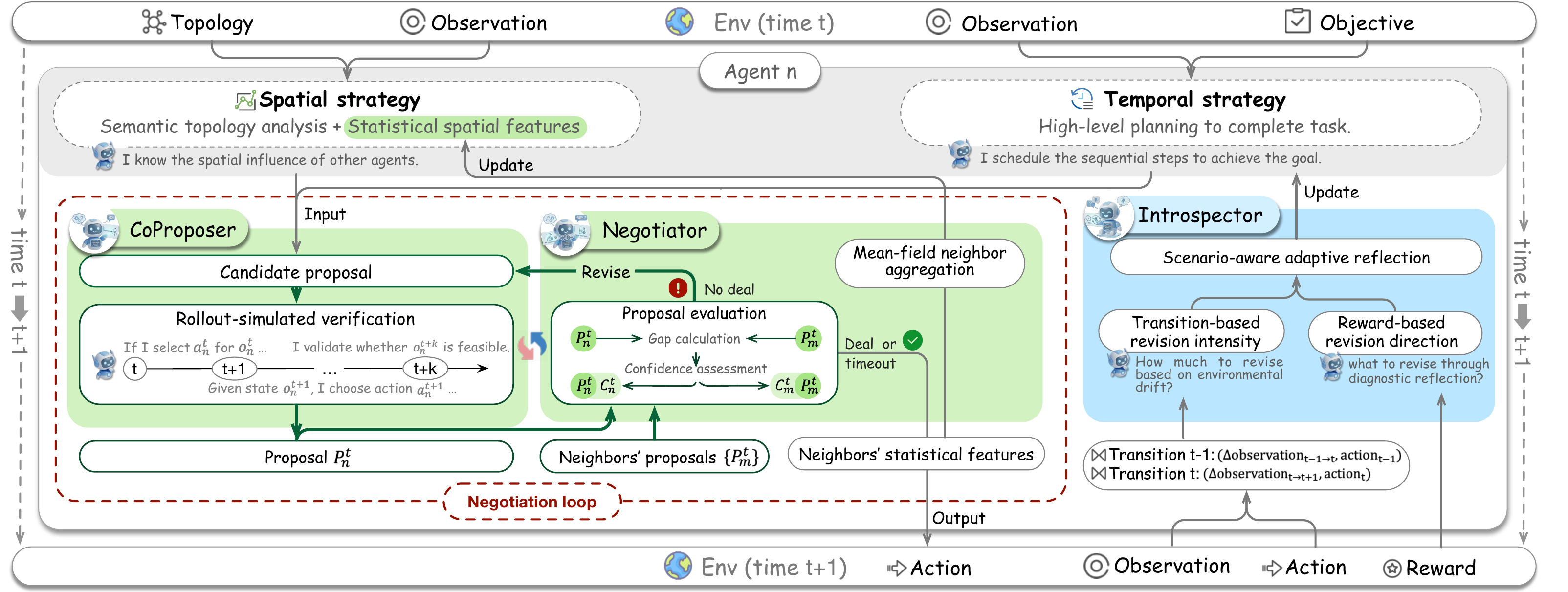}
    \caption{Architecture of MACRO-LLM for agent $n$. The framework comprises three synergistic modules: (1) the CoProposer generates proposals; (2) the Negotiator handles conflict resolution and spatial strategy updates; and (3) the Introspector performs strategy refinement. The CoProposer and Negotiator form a negotiation loop to process observations at time $t$ and output coordinated actions, leading to the next observation at $t+1$.}
    \label{fig:framework}
\end{figure*}

MACRO-LLM establishes a decentralized negotiation framework for MAS under spatiotemporal partial observability. 
As shown in Fig.~\ref{fig:framework}, the architecture of agent $n$ comprises three synergistic modules: \textbf{(1) CoProposer}, for collaborative proposal generation; \textbf{(2) Negotiator}, for conflict resolution and spatial strategy updates; and \textbf{(3) Introspector}, for temporal strategy refinement via scenario-aware adaptive reflection.

\subsection{Problem Setting and System Overview}

\paragraph{Problem Setting.}
We consider strategic-level coordination in a cooperative multi-agent system whose peer-to-peer communication structure is represented by a graph $\mathcal{G}=(\mathcal{V},\mathcal{E})$. Here, $\mathcal{V}$ is the set of agents, and $\mathcal{E}$ specifies the available communication links. The neighborhood of agent $n$ is $\mathcal{N}_n=\{m\in\mathcal{V}\mid(n,m)\in\mathcal{E}\}$. We assume that $\mathcal{G}$ is connected during each negotiation cycle and that agents communicate only with their immediate neighbors.

At step $t$, agent $n$ observes only $o_n^t=\Omega_n(\mathbf{s}^t)$, where $\mathbf{s}^t$ is the unobserved global state. All agents share a semantic team goal $\mathcal{J}_{\mathrm{global}}$ that specifies the task objective and relevant constraints. They use semantic domain knowledge to assess the short-horizon feasibility of their decisions. During each negotiation round, an agent reasons only from its local observation, bounded interaction history $\mathcal{H}_n^t$, current spatial and temporal strategies, and the latest proposals from its neighbors.

These information constraints give rise to coupled spatial and temporal partial observability. Spatially, the local observation mapping $\Omega_n$ and graph-restricted communication limit each agent to a subset of the global state and messages from its immediate neighbors. Temporally, the bounded interaction history provides limited evidence for anticipating future states and actions. Agents must coordinate with incomplete information about both non-local agents and future system evolution.

\paragraph{System Overview.}
Within this setting, MACRO-LLM equips each agent with three modules, as illustrated in Fig.~\ref{fig:framework}. The \textbf{CoProposer} generates and verifies collaborative proposals, the \textbf{Negotiator} resolves proposal conflicts and updates the spatial strategy, and the \textbf{Introspector} uses post-execution feedback to refine the temporal strategy. The CoProposer and Negotiator integrate spatial and temporal reasoning within each environment step. Estimating future action effects requires information about neighborhood states, while resolving spatial conflicts requires reasoning about their temporal consequences. The Introspector provides feedback across environment steps to guide subsequent negotiation cycles.

\subsection{CoProposer}
\label{sec:macro-1}
The CoProposer generates collaborative proposals $P_n^t$ comprising actions for agent $n$ and its observable neighbors $\mathcal{N}_n$ by integrating long-term objectives with immediate spatial constraints.

\paragraph{Temporal Strategy Initialization.}
At the beginning of environment step $t$, agent $n$ generates its temporal strategy $\Pi_{n,t}^{\mathrm{temp}}$ based on its current observation $o_n^t$, action space $\mathcal{A}_n$, and global objective $\mathcal{J}^{global}$, i.e., $\Pi_{n,t}^{\mathrm{temp}} = \operatorname{LLM} \left(o_n^t,\mathcal{A}_n,\mathcal{J}^{\mathrm{global}}\right)$. $\Pi_{n,t}^{\mathrm{temp}}$ provides a high-level plan for progressing toward the global objective.

\paragraph{Spatial Strategy Initialization.} 
At the beginning of $t$, agent $n$ initializes its spatial strategy $\Pi_{n,t}^{\mathrm{spatial},0}$ based on its local observation and the available topological information in $\mathcal{G}$. $\Pi_{n,t}^{\mathrm{spatial},0}$ provides a topological prior for reasoning about how the subsequent actions of neighboring agents may affect local coordination. During negotiation, the Negotiator refines this initial strategy using mean-field spatial features, yielding $\Pi_{n,t}^{\mathrm{spatial},\ell}$ at round $\ell$.


\paragraph{Proposal Generation via Rollout-Simulated Verification.}
Given its current information $\mathcal{I}_n^{t,0}$ and its temporal and spatial strategies, agent $n$ generates an initial semantic proposal
$P_n^{t,0} = (o_n^t, a_n^{t,0}, \{a_{n\rightarrow m}^{t,0}\})$
where $a_n^{t,0}\in\mathcal{A}_n$ is agent $n$'s candidate action,
and $a_{n\rightarrow m}^{t,0}$ is its action recommendation for
neighbor $m$. Each agent ultimately determines its own action.

Because the consequences of a proposal are uncertain, agent $n$ evaluates each proposal through an LLM-generated rollout. Conditioned on the proposal and starting from $o_n^t$, the LLM recursively predicts a $k$-step sequence of observations and actions using domain knowledge, the temporal strategy $\Pi_{n,t}^{\mathrm{temp}}$, and the initial spatial strategy $\Pi_{n,t}^{\mathrm{spatial},0}$. This process requires no external model. The resulting trajectory is defined below and used to assess proposal feasibility and expected improvement.

\noindent\textbf{Rollout}: \textit{In our framework, a rollout is a multi-step LLM-based projection from time $t$ to $t+k$ that infers agent $n$'s future states and actions under its current strategy. The sequence of predicted observations, actions, and rewards across these $k$ steps forms a trajectory, denoted as $\tau_n = \Big( (o_n^t, a_n^t, R_n^t), \dots, (o_n^{t+k}, a_n^{t+k}, R_n^{t+k}) \Big)$.}

To evaluate a rollout, we employ a progressive constraint relaxation mechanism. For the immediate step ($t$), the action must strictly satisfy all environmental constraints to ensure safety. For subsequent steps ($t+1$ to $t+k$), considering increasing uncertainty, criteria are relaxed to minimum viability constraints. This ensures the action $a_n^t$ is safe while conducive to future positive outcomes. The estimated cumulative reward is formulated as:
\begin{equation} \label{equ:Q_func}
\mathcal{R}(\tau_n) = \sum_{\iota=0}^{k} \gamma^\iota R_n(o_n^{t+\iota}, a_n^{t+\iota}).
\end{equation}

To reduce computational complexity over long horizons, estimated per-step scores beyond $t+1$ are replaced by a binary feasibility indicator ($r=1$ if constraints are met, else $0$). If a proposal violates constraints during the rollout, agent $n$ revises its proposed action until a valid proposal is found or the maximum attempt limit is reached. The candidate with the highest $\mathcal{R}(\tau_n)$ is selected and broadcast to neighbors.
The pseudocode is provided in \textbf{Appendix} \ref{app:algo}.

\subsection{Negotiator}
\label{sec:macro-2}

In decentralized environments, agents generate proposals concurrently, inevitably introducing conflicts that necessitate a negotiation mechanism.

\paragraph{Neighborhood State Aggregation via Mean-Field Approximation.}
To extend agents' perception beyond observable limits, we complement semantic reasoning with statistical state features. In complex topologies, exchanging raw observations is bandwidth-inefficient. To address this, we introduce a protocol based on mean-field approximation to transmit aggregated statistical features.

The core idea of mean-field theory approximates complex pairwise interactions between an individual and all other agents in the environment by the interaction between the individual and an aggregated “mean field” effect \cite{lasry2007mean}. Accordingly, we represent the neighborhood influence through weighted statistics. 
For each neighbor $m\in\mathcal{N}_n$, let $s_m^t$ denote its state at step $t$. Under the local mean-field approximation, $s_m^t$ is decomposed into a local mean field and a fluctuation term $s_m^t = \mu_n + \delta s_{n\rightarrow m}^{t}$, and
\begin{equation}
\quad\mu_n = \frac{\sum_{m\in \mathcal{N}_n} w_{n,m} s_m^t}{\sum_{m \in \mathcal{N}_n} w_{n,m}},\label{eq:mean_field_observation}
\end{equation}
where $\mu_n$ represents the weighted mean of the neighborhood's spatial features. The weighted variance is defined as:
\begin{equation}
\sigma_n^2 = \frac{\sum_{m \in \mathcal{N}_n} w_{n,m} \|s_m^t - \mu_n\|_2^2}{\sum_{m \in \mathcal{N}_n} w_{n,m}}.\label{eq:mean_field_variance}
\end{equation}
Here, weights $w_{n,m}>0$ are determined by domain-specific relational factors, such as the spatial distance between agents. During negotiation, agents transmit these statistical features instead of raw data to characterize unobservable contexts. Once agent $n$ receives statistical features $(\mu_m, \sigma_m^2, W_m)$ from a neighbor $m$, it updates the aggregated features incorporating its own state $s_n^t$ (with weight $w_n$) using the weighted Welford algorithm \cite{welford1962note}. {{Since each neighbor already aggregates its own neighborhood's states into the transmitted statistics, this enables agent $n$ to incorporate spatial insights without requiring access to the raw states of all agents. The update equation is provided in \textbf{Appendix} \ref{app:welford}.}}

Eq. \ref{equ:Q_func} defines the estimated cumulative reward $\mathcal{R}(\tau_n)$ for agent $n$'s rollout trajectory $\tau_n$. The trajectory $\tau_n$ depends on the neighborhood states $\mathbf{s}_{\mathcal{N}_n}^t = (s_1^t, \ldots, s_{|\mathcal{N}_n|}^t)$, which influence the agent's spatial strategy $\Pi_{n,t}^{(\text{spatial},\ell)}$. 
Let
$\widehat{R}_n(\mathbf{s}_{\mathcal N_n}^t)
\triangleq R(\tau_n)$
denote the LLM-estimated rollout score.
$\mathcal{R}_n(\mathbf{s}_{\mathcal N_n}^t)$
denote the domain true return used in the following
theoretical analysis.
Therefore, the error is denoted as $\left| \mathcal{R}_n(\mathbf{s}_{\mathcal{N}_n}^t) - \mathcal{R}_n(\boldsymbol{\mu}_n) \right|$
i.e., the gap between the rollout reward computed with full neighborhood information $\mathbf{s}_{\mathcal{N}_n}^t$ and with the mean-field approximation vector $\boldsymbol{\mu}_n = (\mu_n, \ldots, \mu_n)$.

Following the analysis of Yang et al.~\cite{yang2018mean}, we use $\sigma_n^2$ as an indicator of mean-field approximation quality.
In their Eqs.~(7)(8) and Appendix B, pairwise Q-function interactions are approximated by the interaction with a mean agent, with the approximation error controlled by the variance of deviations from the mean~\cite{yang2018mean}. We apply the principle to analyze the theoretical properties of our mean-field aggregation, yielding the error bound:
\begin{equation}\label{eq:thm1}
\left| \mathcal{R}_n(s^t_{\mathcal{N}_n}) - \mathcal{R}_n(\mu_n) 
\right| \leq \frac{L_2}{2w_{\min}} \cdot \sigma^2_n,
\end{equation}
where $L_2$ is a smoothness constant determined by the continuity of the underlying domain dynamics and the boundedness of the state-action spaces, which hold broadly in physically grounded environments with continuous dynamics and finite horizons (a detailed discussion is provided in \textbf{Appendix}~\ref{app:scope}). $\sigma^2_n$ controls the approximation error bound. In many topologies, agent $n$ is a cut vertex connecting disjoint subgroups (e.g., predecessor and follower in the CPP linear topology). Since these subgroups lack direct communication, their state distributions naturally diverge, making a single $\mu_n$ a poor summary. 
We consider a disjoint partition $\{\mathcal{N}_n^{(p)}\}_{p=1}^{P}$ of $\mathcal{N}_n$. For each subgroup $p$, let $W^{(p)}=\sum_{m\in\mathcal{N}_n^{(p)}}w_{n,m}$ and let $\mu_n^{(p)}$ and $\sigma_n^{2,(p)}$ denote the weighted mean and variance computed using the renormalized subgroup weights $w_{n,m}/W^{(p)}$. Let $\boldsymbol{\mu}_n^{(1:P)}$ denote the blockwise-constant vector whose component corresponding to $m\in\mathcal{N}_n^{(p)}$ is $\mu_n^{(p)}$. We normalize the positive weights without loss of generality, such that $\sum_{m\in\mathcal{N}_n}w_{n,m}=1$ and $w_{\min}:=\min_{m\in\mathcal{N}_n}w_{n,m}>0$. Following Appendix~\ref{app:scope}, we further assume that $\nabla \mathcal{R}_n\bigl(\boldsymbol{\mu}_n^{(1:P)}\bigr)^\top\left(\mathbf{s}_{\mathcal{N}_n}^{t}-\boldsymbol{\mu}_n^{(1:P)}\right)=0$.

\begin{theorem}[Tighter Bound via Topology-Aware Partitioning] \label{thm:tighter_bound}
Under the assumptions in \textbf{Appendix}~\ref{app:scope} and general error bound detailed in Eq. \ref{eq:thm1},
computing separate weighted statistics per subgroup yields a tighter bound 
$
\left| \mathcal{R}_n(\mathbf{s}_{\mathcal{N}_n}^{t})
- \mathcal{R}_n(\boldsymbol{\mu}_n^{(1:P)}) \right|
\leq
\frac{L_2}{2\,w_{\min}}
\sum_{p=1}^{P} W^{(p)}\sigma_n^{2,(p)}
\leq
\frac{L_2}{2\,w_{\min}}\sigma_n^2
$.
\end{theorem}
The variance inequality is strict whenever
$\sum_{p=1}^{P}W^{(p)}(\mu_n^{(p)}-\mu_n)^2>0$.
The proof is provided in \textbf{Appendix}~\ref{app:tighter_bound}.
Both bounds are controlled by $\sigma_n^2$. Thus $\sigma_n^2$ serves a dual role: (1) \textit{error indicator}: a smaller $\sigma_n^2$ reflects                                                                                                                                                                                                                                                                                                                                                                                                                                                                                                                                                                                                                                                                                                                                                                                                                                                                                                                                                                                                                                                                                                                                                                                                                                                                                                                                                                                                                                                                                                                                                                                                                                                                                                                                                                                                                                                                                                                                                                                                                                                                                                                                                                                                                                                                                                                                                                                                                                                                                                                                                                                                                                                                                                                                                                                                                                                                                                                                                                                                                                                                                                                                                                                                                                                                                                                                                                                                                                                                                                                                                                                                                                                                                                                                                                                                                                                                                                                                                                                                                                                                                                                                                                                                                                                                                                                                                                                                                                                                                                                                                                                                                                                                                                                                                                                                                                                                                                                                                                                                                                                                                                                                                                                                                                                                                                                                                                                                                                                                                                                                                                                                                                                                                                                                                                                                                                                                                                                                                                                                                                                                                                                                                                                                                                                                                                                                                                                                                                                                                                                                                                                                                                                                                                                                                                                                                                                                                                                                                                                                                                                                                                                                                                                                                                                                                                                                                                                                                                                                                                                                                                                                                                                                                                                                                                                                                                                                                                                                                                                                                                                                                                                                                                                                                                                                                                                                                                                                                                                                                                                                                                                                                                                                                                                                                                                                                                                                                                                                                                                                                                                                                                                                                                                                                                                                                                                                                                                                                                                                                                                                                                                                                                                                                                                                                                                                                                                                                                                                                                                                                                                                                                                                                                                                                                                                                                                                                                                                                                                                                                                                                                                                                                                                                                                                                                                                                                                                                                                                                                                                                                                                                                                                                                                                                                                                                                                                                                                                                                                                                                                                                                                                                                                                                                                                                                                                                                                                                                                                                                                                                                                                                                                                                                                                                                                                                                                                                                                                                                                                                                  a more accurate mean-field approximation; (2) \textit{uncertainty signal}: large $\sigma_n^2$ prompts conservative strategies. During negotiation, agents progressively align their strategies, reducing $\sigma_n^2$ and tightening both bounds over successive rounds.

\paragraph{Proposal Confidence Assessment and Regeneration.}
To evaluate proposal reliability, agent $n$ computes the pairwise differences between its own proposal $P_n^t$ and received neighbor proposals $\{P_m^t\}_{m \in \mathcal{N}_n}$. 
Local consensus between agents $n$ and $m$ is defined as $|P_n^{(q)} - P_m^{(q)}| < \delta$, where $\delta = \epsilon \cdot (a_{\max} - a_{\min})$ is proportional to the action space range for continuous actions, and $\delta=0$ for discrete action spaces requiring exact agreement. If consensus is not met, the agent analyzes discrepancies, updates its spatial strategy by incorporating the received mean-field states, and integrates weighted neighbor proposals to regenerate a refined candidate, which then undergoes rollout-simulated verification as detailed in \textbf{Sec. \ref{sec:macro-1}}.

\paragraph{Multi-Round Negotiation.}
A complete CoProposer--Negotiator iteration (\textbf{Secs.~\ref{sec:macro-1}} and \textbf{\ref{sec:macro-2}}) constitutes a 
\emph{negotiation round}. Multi-round negotiation primarily propagates information: each round extends agents' effective perception by one communication hop, so $r$ rounds cover $r+1$ hops. Local consensus serves as an early stopping criterion. Upon completion, agent $n$ produces a final decision by weighting each neighbor's proposal by the confidence scores from proposal evaluation, selecting the highest-confidence action (or weighted combination for continuous actions). Proposition~\ref{thm:nego} characterizes bounded termination and shows that proposal quality does not decrease across negotiation rounds under the idealized assumption of fully reliable LLM inference.

\begin{proposition}[Properties of the Negotiation Loop]
\label{thm:nego}
Let $\mathcal{G}$ be a connected graph and let $r$ denote the maximum negotiation rounds. Under an idealized rational-inference setting in which the LLM ranks candidate proposals consistently with their rollout quality, the negotiation loop satisfies the following properties. (1) Bounded Termination: The negotiation terminates within $r$ rounds, either via early stopping when local consensus is reached ($\|P_n - P_m\| < \delta,\ \forall m \in \mathcal{N}_n$) or via the maximum round limit. (2) Monotonic Non-Decrease of Proposal Quality: $\mathcal{R}(\tau_n^{(\ell+1)}) \geq \mathcal{R}(\tau_n^{(\ell)}), \quad \forall\, \ell = 0, 1, \ldots, r-1$.
\end{proposition}
The proof can be found in \textbf{Appendix} \ref{app:nego}. Regardless of whether agents reach consensus, the negotiation process terminates within $r$ rounds, and the retained best estimated rollout score is non-decreasing across rounds by construction.

\subsection{Introspector}
\label{sec:macro-3}
Under spatiotemporal partial observability, state inference becomes less reliable because of environmental changes and uncertainty about other agents' behavior. A reward decline between consecutive time steps triggers self-reflection, through which the agent revises its strategy.
We model this process as \textit{scenario-aware adaptive reflection}, which determines \textit{how much} to revise based on environmental drift and \textit{what} to revise through diagnostic attribution.

\paragraph{Revision Intensity Estimation.}
We first quantify environmental shifts to determine the revision intensity. Let $o_n^t$ denote the vectorized observation of agent $n$ at time $t$. We construct a transition vector $\Gamma_n^t$ by concatenating the min-max normalized observation difference and the action vector, defined as $\Gamma_n^t = (\Delta{\tilde{o}_n^t}, \tilde{a}_n^t)$, where $\Delta{o_n^t} = o_n^{t+1}-o_n^t$.

To evaluate scene consistency, we compute the cosine similarity between two consecutive transitions, $\Gamma_n^{t-1}$ and $\Gamma_n^t$. The drift intensity ($\lambda_{\text{drift}}$) is derived from scene discontinuity:
\begin{equation}
\lambda_{\text{drift}} = 1 - \frac{\Gamma_n^{t-1}\cdot \Gamma_n^{t}}{\|\Gamma_n^{t-1}\| \cdot \|\Gamma_n^{t}\|}.
\label{eq:drift_intensity}
\end{equation}
Higher $\lambda_{\text{drift}}$ indicates greater environmental drift requiring aggressive revision, while lower values permit only fine-grained refinements. Together with the reward-decline trigger, this confines revisions to genuine degradation.

\paragraph{Revision Direction Identification.}
We generate a \textit{semantic revision signal} as a natural language description of the ``direction'' for strategic improvement. The LLM analyzes observation logs and negotiation records $\mathbf{N}_t$ to identify which transitions deviated from the temporal strategy and to assess the impacts from neighbors. The revision signal $g_{n,t}^{\text{semantic}}$ is generated as textual rules that characterize the associations between actions, state changes, and negative reward outcomes, providing diagnostic guidance for strategy refinement.

\paragraph{Strategy Update.}
Finally, we update the temporal strategy using the revision signal and drift intensity. A concrete prompt template for strategy update with revision intensity and direction is provided in \textbf{Appendix}~\ref{app:prompt_intro}. The update is formulated as $\Pi_{n,t+1}^{\text{temp}} \leftarrow \text{LLM}\left(\Pi_{n,t}^{\text{temp}},\ g_{n,t}^{\text{semantic}},\ \lambda_{\text{drift}}\right)$. 
The drift intensity $\lambda_{\mathrm{drift}}$ controls the extent of the revision. A high value indicates substantial environmental drift and prompts the LLM to revise the strategy according to $g_{n,t}^{\mathrm{semantic}}$. A low value limits the revision to fine-grained adjustments while preserving the core of $\Pi_{n,t}^{\mathrm{temp}}$.
\section{Experiments}

\subsection{Experimental Settings}
\begin{table}[t!]
\centering
\caption{Performance comparison on the CPP task across \textit{Catch-up} and \textit{Slow-down} scenarios. Metrics include RMSE and SD for Headway (H) and Velocity (V). \textbf{Bold} and \underline{underline} indicate best and second-best results, respectively.}
\label{tab:cacc}
\resizebox{\linewidth}{!}{
\begin{tabular}{lcccc}
\toprule
\multicolumn{5}{c}{\cellcolor{gray!10}\textbf{Scenario 1: Catch-up}} \\
\midrule
\textbf{Methods} & RMSE-H ($\downarrow$) & RMSE-V ($\downarrow$) & SD-H ($\downarrow$) & SD-V ($\downarrow$) \\
\midrule
ToM-Belief & 5.514 & 4.409 & 4.002 & 3.130 \\
ChatEval   & 5.495 & 2.113 & 2.947 & 1.241 \\
LAMEN      & 1.891 & 0.648 & 3.320 & 0.601 \\
\midrule
DPPO       & 1.603 & 0.978 & 0.994 & 0.474 \\
DMPO       & 4.765 & \textbf{0.534} & 3.025 & 1.349 \\
IC3Net     & 8.687 & 4.205 & 7.232 & 3.399 \\
\midrule
\rowcolor[HTML]{E6F4EA}
MACRO-DS$^\dagger$ & 1.528 & \underline{0.553} & \underline{0.611} & \textbf{0.190} \\
\rowcolor[HTML]{E6F4EA}
MACRO-Llama$^\dagger$ & \underline{1.412} & 0.779 & 0.645 & 0.365 \\
\rowcolor[HTML]{E6F4EA}
MACRO-GPT$^\dagger$ & \textbf{1.237} & 0.721 & \textbf{0.409} & \underline{0.203} \\
\midrule
\midrule
\multicolumn{5}{c}{\cellcolor{gray!10}\textbf{Scenario 2: Slow-down}} \\
\midrule
\textbf{Methods} & RMSE-H ($\downarrow$) & RMSE-V ($\downarrow$) & SD-H ($\downarrow$) & SD-V ($\downarrow$) \\
\midrule
ToM-Belief & \underline{0.641} & 4.681 & \underline{0.225} & 0.480 \\
ChatEval   & 3.062 & 5.708 & 0.778 & 1.016 \\
LAMEN      & 0.923 & 3.385 & 0.635 & 0.750 \\
\midrule
DPPO       & 1.720 & \underline{3.121} & 1.410 & 1.176 \\
DMPO       & 2.226 & \textbf{3.059} & 1.742 & 1.189 \\
IC3Net     & 7.877 & 4.678 & 5.207 & 2.319 \\
\midrule
\rowcolor[HTML]{E6F4EA}
MACRO-DS$^\dagger$ & 0.649 & 3.388 & 0.274 & \underline{0.417}\\
\rowcolor[HTML]{E6F4EA}
MACRO-Llama$^\dagger$ & 1.163 & 3.736 & 0.889 & 1.224 \\
\rowcolor[HTML]{E6F4EA}
MACRO-GPT$^\dagger$ & \textbf{0.496} & 3.310 & \textbf{0.128} & \textbf{0.238} \\
\bottomrule
\multicolumn{5}{l}{$^\dagger$ MACRO-\texttt{X}: DS denotes DeepSeek-V3.1, Llama denotes} \\
\multicolumn{5}{l}{Llama-3.3-70B-Instruct, and GPT denotes GPT-4o.} \\
\end{tabular}
}
\end{table}

\paragraph{Task Configurations.}
We evaluate MACRO-LLM across two complementary tasks where coordination outcomes are measured through task-level environment metrics, ensuring that performance gains reflect collaborative decision quality rather than language fluency.
(1) \textit{Cooperative Platoon Planning (CPP)~\cite{chu2020multi}}: a continuous-dynamics coordination benchmark for vehicle platooning (default N=8) with dynamic ``Catch-up'' and ``Slow-down'' scenarios.
(2) \textit{Pandemic Control (PC)~\cite{kompella2020reinforcement}}: a strategic planning task where agents select regulation level to mitigate virus spread across three urban topologies (Helsinki, Hong Kong, and New York).
CPP and PC tasks span continuous and discrete action spaces, linear and mixed topologies, and fine-grained and long-horizon decision-making, respectively. Both tasks are built on widely-used, physically-grounded simulators.
Refer to \textbf{Appendix}~\ref{app:env_detail} for configurations and city topologies. 

\paragraph{Implementation Details.}
We implement MACRO-LLM as a backbone-agnostic framework and evaluate it across six LLM backbones: GPT-4o, GPT-4o-mini, DeepSeek-V3.1, Llama-3.3-70B-Instruct, Mistral-Small-3.2-24B-Instruct, and Qwen3-flash. Tables~\ref{tab:cacc} and~\ref{tab:pan} report three representative backbones. The full backbone robustness analysis is in \textbf{Appendix}~\ref{app:gen}. Scalability, ablation, and robustness analyses use GPT-4o as the default. Framework and task configurations are detailed in \textbf{Appendix}~\ref{app:env_detail}. MACRO-LLM results are averaged over five independent runs. The full variance analysis is provided in \textbf{Appendix}~\ref{app:robust}. A case study on hierarchical deployment with different LLM call frequency is in \textbf{Appendix}~\ref{app:deploy}.

\begin{table*}[!t]
\centering
\renewcommand{\arraystretch}{1}
\caption{Performance comparison on the PC task across Helsinki, Hong Kong, and New York topologies. Metrics include Normalized Infection (I\_n), Peak Infection (PI\_n), Normalized Death (D\_n), and Pandemic Duration (PD). \textbf{Bold} and \underline{underline} indicate best and second-best results, respectively.}
\label{tab:pan}
\resizebox{0.95\textwidth}{!}
{\begin{tabular}{lllllllllllll}
\toprule
\multirow{2}{*}{\textbf{Methods}} 
 & \multicolumn{4}{c}{\textbf{Helsinki}} 
 & \multicolumn{4}{c}{\textbf{Hong Kong}} 
 & \multicolumn{4}{c}{\textbf{New York}} \\
\cmidrule(lr){2-5} \cmidrule(lr){6-9} \cmidrule(lr){10-13}
 & I\_n ($\downarrow$) & PI\_n ($\downarrow$) & D\_n ($\downarrow$) & PD ($\downarrow$)
 & I\_n ($\downarrow$) & PI\_n ($\downarrow$) & D\_n ($\downarrow$) & PD ($\downarrow$) 
 & I\_n ($\downarrow$) & PI\_n ($\downarrow$) & D\_n ($\downarrow$) & PD ($\downarrow$) \\
\midrule
ToM-Belief & 0.184 & 0.042 & 0.002 & 92 
& 0.239 & 0.028 & 0.003 & 120 
& 0.130 & 0.023 & \underline{0.001} & 60 \\ 
ChatEval & \textbf{0.010} & \textbf{0.008} & \textbf{0.000} & 42 
& \underline{0.009} & \textbf{0.003} & \underline{0.001} & 24 
& 0.290 & 0.043 & 0.002 & 83\\
LAMEN & \textbf{0.010} & \textbf{0.008} & 0.002 & 17 
& 0.012 & 0.005 & \textbf{0.000} & 23 
& 0.164 & 0.025 & 0.002 & 52 \\
\midrule
DPPO & 0.426 & 0.110 & 0.007 & 62 
& 0.370 & 0.083 & 0.004 & 83 
& 0.923 & 0.363 & 0.009 & 95\\
DMPO & \textbf{0.010} & \textbf{0.008} & \underline{0.001} & \underline{13} 
& 0.271 & 0.057 & 0.003 & 59 
& 0.332 & 0.109 & 0.002 & 53 \\
IC3Net & 0.651 & 0.186 & 0.009 & 69 
& 0.259 & 0.067 & 0.003 & 52 
& 0.789 & 0.284 & 0.007 & 93 \\
\midrule

\rowcolor[HTML]{E6F4EA}
MACRO-DS$^\dagger$ & 0.098 & \underline{0.038} & 0.002 & 46
& \textbf{0.008} & 0.005 & \underline{0.001} & \underline{22}
& \textbf{0.003} & \textbf{0.003} & \textbf{0.000} & \textbf{11}
\\
\rowcolor[HTML]{E6F4EA}
MACRO-Llama$^\dagger$ & 0.013 & \textbf{0.008} & \textbf{0.000} & 14
& 0.012 & 0.005 & \textbf{0.000} & \textbf{16}
& 0.013 & \underline{0.006} & \textbf{0.000} & \underline{25}
\\
\rowcolor[HTML]{E6F4EA}
MACRO-GPT$^\dagger$ & \underline{0.011} & \textbf{0.008} & \textbf{0.000} & \textbf{10}
& \textbf{0.008} & \underline{0.004} & \underline{0.001} & 23
& \underline{0.004} & \textbf{0.003} & \textbf{0.000} & \underline{25}
\\
\bottomrule
\multicolumn{13}{l}{$^\dagger$ MACRO-\texttt{X}: DS denotes DeepSeek-V3.1, Llama denotes Llama-3.3-70B-Instruct, and GPT denotes GPT-4o.} \\
\end{tabular}}
\end{table*}

\paragraph{Baselines.}
We benchmark MACRO-LLM against representative methods from two paradigms: (1)~\textit{LLM-based MAS baselines}, which share the same backbone (GPT-4o) and environmental interfaces to provide controlled comparisons: ToM-Belief~\cite{li2023theory}, ChatEval~\cite{chan2023chateval}, and LAMEN~\cite{davidson2024evaluating}; and (2)~\textit{MARL baselines}, which serve as task-specialized performance references trained on each topology: DPPO, DMPO~\cite{ma2024efficient}, and IC3Net~\cite{ic3net2018learning}. Details on baseline implementations, prompt adaptations, and the division between LLM reasoning and external numerical modules are provided in \textbf{Appendix}~\ref{app:system_design}.

\paragraph{Evaluation Metrics.}
For the CPP task, we evaluate platoon stability via the Root Mean Square Error (\textit{RMSE}) and Standard Deviation (\textit{SD}) of both velocity and headway distance (denoted as RMSE-V/H and SD-V/H). For the PC task, we assess containment efficacy through Normalized Infection (\textit{I\_n}), Peak Infection (\textit{PI\_n}), Normalized Death (\textit{D\_n}) and total Pandemic Duration (\textit{PD}).
Metric definitions are detailed in \textbf{Appendix}~\ref{app:metrics}.

\subsection{Performance of Cooperative Platoon Planning Task}
\label{experimentresults}

As shown in Table~\ref{tab:cacc}, MACRO-LLM generally outperforms both MARL and LLM-based baselines across multiple backbones. In the \textit{Catch-up} scenario, all three MACRO-LLM variants achieve the top RMSE-H and SD-H results. The SD-H of MACRO-GPT ($0.409 \pm 0.034$, \textbf{Appendix} \ref{app:robust}) is 87.7\% lower than that of LAMEN and 58.9\% lower than that of DPPO, while even MACRO-DS attains lower SD-H than all baselines. In the \textit{Slow-down} scenario, MACRO-LLM prioritizes stable headway coordination over velocity matching. MACRO-GPT achieves the lowest SD-H ($0.128 \pm 0.054$, \textbf{Appendix} \ref{app:robust}), a 43.1\% improvement over ToM-Belief, while maintaining competitive velocity errors. The consistent advantage across three diverse backbones confirms that the performance gains stem from the framework's negotiation mechanism rather than any single model's capability.

\begin{figure}[!t]
    \centering
    \includegraphics[width=0.9\linewidth]{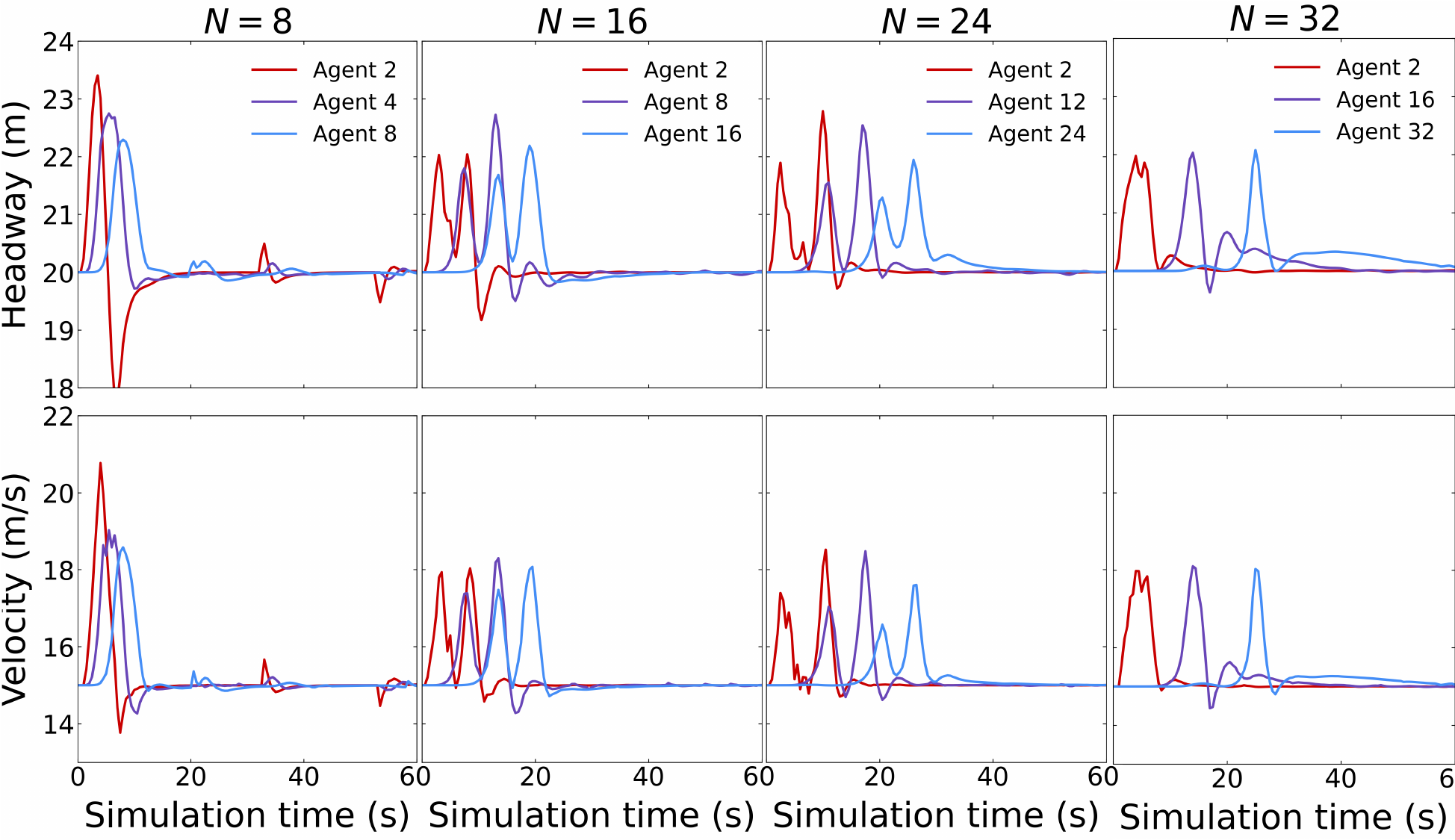}
    \caption{Scalability analysis of MACRO-LLM on the CPP task for number of agents $N \in \{8, 16, 24, 32\}$. }
    \label{fig:scalability}
\end{figure}

\subsection{Performance on Pandemic Control Task}

Table~\ref{tab:pan} presents the comparative results across three city topologies. MACRO-LLM achieves the shortest pandemic duration with at least one backbone, while generally maintaining low infection rates. In \textit{New York} topology, 
all three MACRO-LLM variants achieve substantially lower cumulative infections than DPPO and shorten the pandemic duration from 95 to as few as 11 days (MACRO-DS).
In \textit{Hong Kong}, MACRO-Llama reduces PD to 16 days, outperforming LAMEN (23 days) at comparable infection levels. MACRO-LLM achieves both low infections and short durations simultaneously. This consistent advantage across three diverse backbones confirms that the coordination gains stem from the framework's negotiation mechanism rather than any single model's capability.

\subsection{Analysis of Scalability}

To investigate scalability, we extend the CPP Catch-up scenario to fleet sizes $N \in \{8, 16, 24, 32\}$. As illustrated in Fig.~\ref{fig:scalability}, MACRO-LLM maintains consistent robustness as the network quadruples. Oscillation amplitudes remain bounded and all agents converge to the target equilibrium within 40s regardless of fleet size. Aggregated metrics (\textbf{Appendix}~\ref{app:scala}, Table~\ref{tab:scalability}) reveal two complementary trends: RMSE-H decreases from 1.237 ($N$=8) to 0.728 ($N$=32), as the growing proportion of already-stabilized agents dilutes the transient deviations of those still responding to propagation. Meanwhile, SD-H increases moderately from 0.409 to 0.433, reflecting the inherently longer synchronization lag in larger platoons. Importantly, this increase is bounded. At $N$=32, SD-H remains 56.4\% lower than that of DPPO ($N$=8), confirming that MACRO-LLM compresses large-scale interactions while ensuring coordination quality. Further scalability analyses are provided in \textbf{Appendix}~\ref{app:scala}.

\subsection{Ablation Study}
\begin{figure}[!t]
    \centering
    \includegraphics[width=0.9\linewidth]{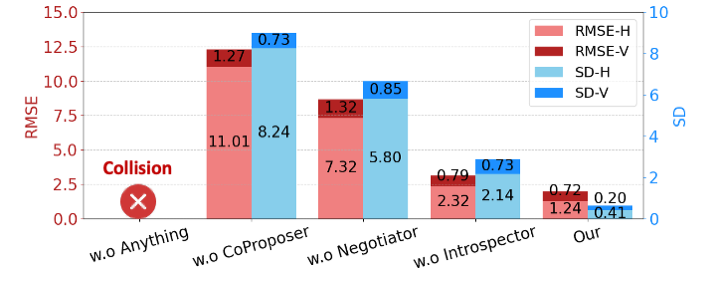}
    \caption{Ablation of MACRO-LLM—impact of each module on CPP performance.}
    \label{fig:exp-abla1}
\end{figure}
\begin{figure}[!t]
    \centering
    \includegraphics[width=0.9\linewidth]{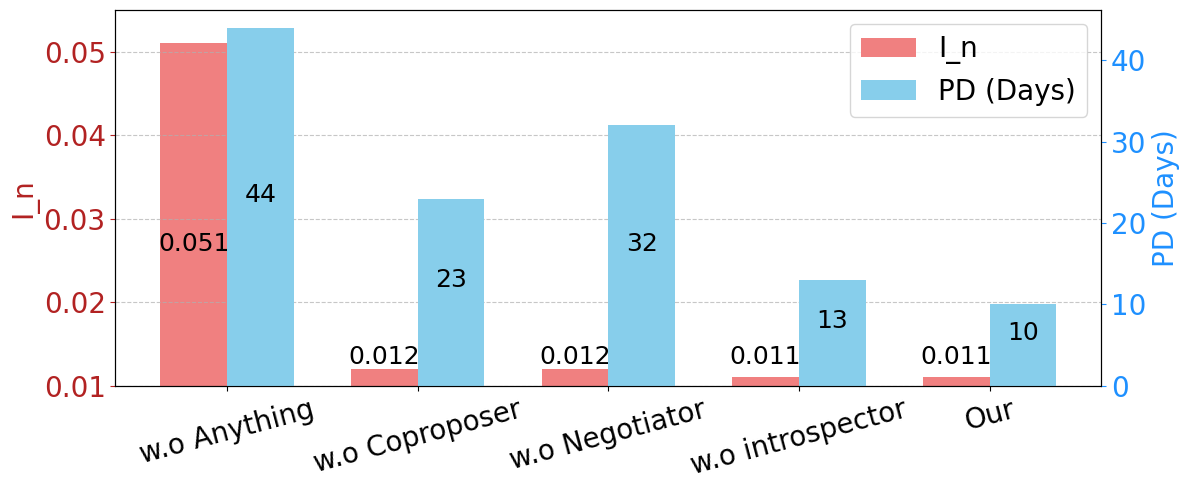}
    \caption{Ablation of MACRO-LLM—impact of each module on PC performance.}
    \label{fig:exp-abla2}
\end{figure}

We validate each module's contribution in Figs.~\ref{fig:exp-abla1} and~\ref{fig:exp-abla2}. Ablating the Negotiator degrades stability metrics drastically. Without the Introspector, agents fail to suppress oscillations over time. Notably, the ablations reveal cross-module dependencies: removing the CoProposer disables not only temporal verification but also the Negotiator's confidence assessment without rollout rewards. This explains the largest SD-H degradation among all variants. Quantitative analyses are in \textbf{Appendix}~\ref{app:ablation}.

\section{Conclusion}
In this work, we present MACRO-LLM, a novel framework enabling agents to reason and coordinate through localized negotiations under spatiotemporal partial observability. We address spatial and temporal coordination challenges through three interdependent modules: (1) a CoProposer that generates collaborative actions and mitigates future uncertainty via predictive rollouts; (2) a Negotiator that resolves spatial conflicts using mean-field statistical estimation; and (3) an Introspector that supports strategic refinement via scenario-aware adaptive reflection. Extensive experiments on CPP and PC tasks demonstrate MACRO-LLM's capability to align local actions with long-term global objectives, achieving superior collaborative efficiency compared to representative baselines. Ultimately, our work highlights the critical role of explicitly modeling spatiotemporal dynamics in scaling LLM-based MAS.

\section*{Limitations}
While MACRO-LLM demonstrates robust coordination capabilities, we acknowledge two primary limitations. 
First, MACRO-LLM is designed as a strategic-level reasoning framework rather than a high-frequency execution controller.
When invoked at every environment step in our benchmark implementation, runtime is dominated by remote API overhead. The runtime can be mitigated through two composable strategies validated in our experiments:
(1) hierarchical deployment, where MACRO-LLM issues strategic directives at coarse decision intervals while a domain-specific execution layer handles fast state updates.
Empirical results confirm that even with 90\% fewer LLM invocations, coordination quality surpasses both representative baselines reported in Table~\ref{tab:hierarchical}; and (2) local deployment of lightweight models such as Qwen-flash and Mistral-small-3.2-24B-Instruct, which eliminates API latency entirely while maintaining competitive performance (\textbf{Appendix}~\ref{app:gen}). Future work will explore tighter co-design between the strategic layer and domain-specific execution layers to further optimize the reasoning-latency balance. Second, relying on foundation models requires additional context injection for domain adaptation, which increases token consumption. 
Future work will explore online self-evolution mechanisms, such as experience-driven prompt distillation and adaptive strategy caching, to reduce context dependency and improve cross-domain adaptability.
Relatedly, the Introspector's semantic refinement is verified to be stable across the continuously differentiable dynamics of CPP and PC (Appendix \ref{app:robust}, and Appendix \ref{app:gen}). Its behavior in chaotic, high-dimensional physical systems with strong sensitivity to initial conditions remains future work, while its influence is bounded to refinement quality rather than direct control reliability by the downstream rollout and consensus checks. More broadly, our evaluation is conducted in physically-grounded simulators, and results remain consistent across two domains, four fleet sizes, and six backbones. Quantifying the sim-to-real gap on physical hardware, where the hierarchical execution layer (Appendix~\ref{app:deploy}) is designed to absorb sensor and actuation noise, is an important direction for future work.

\section*{Acknowledgments}
This work was partially supported by the UGC General Research Fund no. 17209822 and the Innovation and Technology Commission Fund no. ITS/383/23FP from Hong Kong. 
\bibliography{Refe}

\appendix

\clearpage
\appendix
\section*{Appendix Table of Contents}
\startcontents[appendix]
\printcontents[appendix]{l}{1}{\setcounter{tocdepth}{2}}
\section{Extended Related Works}
\label{app:related_works}
Due to space constraints, we provide a comprehensive review of the literature in this appendix.

\paragraph{Agent Planning with LLMs.}
There has been a recent surge in research on Large Language Model (LLM)-based agent reasoning and planning~\cite{huang2024understanding, yuksekgonul2025optimizing}. Techniques such as Chain-of-Thought (CoT)~\cite{wei2022chain}, ReAct~\cite{yao2023react}, Self-Consistency~\cite{wang2022self}, and Graph-of-Thought~\cite{besta2024graph} enhance single-agent reasoning by decomposing tasks or broadening candidate exploration. Self-Refine~\cite{madaan2023self} introduces iterative self-correction, and MemoryBank~\cite{zhong2024memorybank} augments planning with historical knowledge retrieval. However, these methods are designed for single-agent settings with full or near-complete observability. In multi-agent scenarios under partial observability, the core challenge shifts from individual reasoning quality to inter-agent coordination under fragmented information. That is a problem that single-agent planning enhancements do not address.

\paragraph{Extended Related Work on LLM-based Multi-Agent Systems.}
The emergence of LLMs has revolutionized multi-agent collaboration. Early frameworks such as CAMEL~\cite{Li2023CAMEL} and AutoGen~\cite{Wu2023AutoGen} demonstrated the potential of communicative agents to solve open-ended tasks through dynamic role-playing. Subsequent systems introduced structured enhancements to improve stability and capability. For instance, MetaGPT~\cite{Hong2023MetaGPT} incorporates standardized operating procedures to streamline workflows, while Voyager~\cite{Wang2023Voyager} utilizes an iterative curriculum with persistent code libraries for embodied control. Despite their success in observable environments (e.g., software engineering, Minecraft), these frameworks typically operate under the assumption of global observability or unrestricted communication, rendering them less effective in spatially distributed, bandwidth-constrained real-world scenarios. 
Recent research targeting the information imbalance caused by spatial constraints generally falls into two categories:

\begin{enumerate}
    \item \textit{Hierarchical aggregation}: Works such as FinCon~\cite{yu2024fincon}, MLPO~\cite{estornell2025train}, and HALO~\cite{hou2025halo} employ hierarchical architectures. These systems designate specific agents (e.g., central hubs) to aggregate incomplete information, constructing a pseudo-global view to compensate for limited local perception. However, this centralization creates critical bottlenecks due to its reliance on central nodes and the prohibitive communication overhead incurred as the network scales.
    \item \textit{Individual reasoning enhancement:} A complementary line of research focuses on enhancing individual reasoning capabilities. Theory-of-Mind (ToM) approaches~\cite{li2023theory} enable agents to infer the beliefs and intents of others, while frameworks like LAMEN~\cite{davidson2024evaluating} utilize structured reasoning chains to predict hidden states. While effective for small-scale interactions, these methods often struggle with the \textit{spatiotemporal uncertainty} inherent in complex topologies, where agents must simultaneously reason about spatial neighbors and future dynamics under strict latency constraints. 
    \item \textit{Swarm-based optimization:} Recent works apply particle swarm optimization to LLM systems. Model Swarms~\cite{feng2024model} searches the model weight space over 50 iterations with per-particle validation-set evaluation to produce adapted expert checkpoints. SwarmAgentic~\cite{zhang2025swarmagentic} searches the system configuration space over 10 iterations, jointly optimizing agent functionalities and collaboration structures through failure-driven refinement. While both methods achieve dynamic configuration evolution, this adaptation occurs during an offline 
optimization loop that requires a globally accessible fitness function evaluated over the entire validation set. 
\end{enumerate}
A parallel line of work applies LLM-based negotiation to particular physical domains. For instance, CoLMDriver \cite{liu2025colmdriver} introduces an LLM negotiation pipeline for cooperative autonomous driving, coupling vehicle-to-vehicle communication with 3D perception and waypoint-level trajectory planning. Such systems are tightly integrated with domain-specific perception and control stacks, and their evaluation is necessarily bound to the corresponding full-stack simulators. Unlike application-specific end-to-end pipelines, MACRO-LLM isolates the coordination mechanism and tests a shared three-module architecture in two distinct simulators with continuous versus discrete action spaces and linear versus graph topologies. Consequently, our baselines are selected to share a controlled environmental interface, enabling performance differences to be attributed to coordination mechanisms rather than perception or system-integration choices. MACRO-LLM proposes a fully decentralized negotiation mechanism that leverages mean-field statistics and semantic strategies, effectively balancing local perception with global coordination goals.

\paragraph{Extended Related Work on MARL}
\label{app:related_marl}
To tackle spatial partial observability, the decentralized partially observable Markov decision process serves as the foundational modeling framework \cite{oliehoek2016concise}. Under this model, agents must act based on local histories rather than global states. To mitigate the non-stationarity inherent in decentralized learning, the centralized training with decentralized execution paradigm has become standard. Recent works have further incorporated memory augmentation mechanisms \cite{kia2024memory} and theoretical guarantees \cite{bernstein2005bounded} to approximate global state value functions. However, these methods typically rely on value functions learned for a fixed state space, which limits their transferability across environments.

Scalability remains a critical bottleneck in MARL. Mean-field MARL offers a promising direction by approximating agent-to-agent interactions via the average effect of the neighborhood \cite{yang2018mean}. Subsequent studies have extended this to accommodate agent heterogeneity \cite{ganapathi2020multi, gu2025mean} and robustness \cite{wang2022attention, anand2024mean}. 
Similarly, MACRO-LLM adopts the mean-field principle in its \textit{Negotiator} module to compress spatial information. However, unlike MARL which learns a mean-field policy through extensive episodes, our approach computes mean-field statistical features explicitly as prompts for the LLM, enabling inference-time reasoning without gradient updates.

To improve sample efficiency, advanced policy optimization methods such as Multi-Agent PPO (MAPPO) \cite{schulman2017proximal} and Model-based Decentralized Policy Optimization (DMPO) \cite{ma2024efficient} have been developed. Furthermore, adaptive communication protocols like IC3Net \cite{ic3net2018learning} introduce gating mechanisms to decide when to communicate.
Despite these advances, the fundamental reliance on trial-and-error interaction makes MARL computationally intensive.
As noted by \cite{huh2023multi} and \cite{oroojlooy2023review}, MARL policies often struggle to generalize to unseen topologies or agent numbers, necessitating the retraining costs that our LLM-based framework aims to eliminate.

\section{Detailed Algorithm of Proposal Generation} \label{app:algo}

\begin{algorithm*}[ht]
\caption{Pseudocode of proposal generation via rollout-simulated verification.}
\label{alg:rsv}
\KwIn{Observation $o_n^t$; strategies $\Pi_{n,t}^{\text{temp}}$ and $\Pi_{n,t}^{\mathrm{spatial},0}$; maximum attempts $Att_{\text{max}}$; rollout horizon $K$}
\KwOut{Rollout-screened proposal $P_n^t=(o_n^t,{a_n^t},\{ {a_{n\rightarrow m}^t} \}_{m\in\mathcal N_n})$}
Initialize \textsc{Candidate\_Proposals}$\leftarrow\varnothing$ and $att=0$\;
Initialize candidate action with highest LLM-estimated reward score by $a_n^{t,0} \leftarrow \arg\max_{a}\mathsf{InferAction}(o_n^t, \mathcal{A}_n, \Pi_{n,t}^{\text{temp}}, \Pi_{n,t}^{\mathrm{spatial},0}$ and ensure $R_n^t(o_n^{t},a_n^{t,0})\ge R_n^{t-1}(o_n^{t-1},a_n^{t-1})$\;
\While{$att<Att_{max}$}{
    Initialize rollout trajectory $\tau_n^{t,att}=(o_n^t,a_n^{t,att},R_n^t)$\;
    \ForEach{$m\in \mathcal N_n$}{
    Obtain observable agents' candidate action ${a_{n\rightarrow m}^{t,att}} \leftarrow \mathsf{InferNeighborAction}(o_n^t,{a_n^{t,att}}, \Pi_{n,t}^{\mathrm{spatial},0}$\;}
    Construct candidate proposal $P_n^{t,att}\leftarrow (o_n^t,{a_n^{t,att}},\{{a_{n\rightarrow{m}}^{t,att}}\}_{m\in\mathcal N_n})$\;
    \For{Verified time step $k=1,\cdots,K$ in a rollout}{
    Predict next observation $o_n^{t+k,att}$ with candidate proposal $P_n^{t,att}$\;
    \If{$o_n^{t+k,att}$ violates constraints}{
    Append $P_n^{t,att}$ to \textsc{Candidate\_Proposals}\;
    $att=att+1$\;
    Update candidate action $a_n^{t,att}$ slightly and ensure $R_n^t(o_n^{t},a_n^{t,0})\ge R_n^{t-1}(o_n^{t-1},a_n^{t-1})$\;
    Break\;
    }
    Infer actions $a_{n\rightarrow m}^{t+k,att}$ on time $t+k$\;
    Add $(a_n^{t+k,att}, o_n^{t+k,att}, R_n^{t+k,att})$ to rollout trajectory $\tau_n^{t,att}$\;
    }
    \If{$P_n^{t,{att}}$ has been verified over $k$ time steps}
    {
    \Return Selected $P_n^t\leftarrow P_n^{t,att}$. 
    }
}
Select the proposal with the highest LLM-estimated rollout score from \textsc{Candidate\_Proposals} as $P_n^t$\;
\Return $P_n^t$\;
\end{algorithm*}

In this section, we detail the algorithmic procedure for the CoProposer module. The core mechanism, \textit{Proposal Generation via Rollout-Simulated Verification}, is outlined in Algorithm \ref{alg:rsv}.

This process aims to find an action sequence that maximizes cumulative reward while strictly adhering to safety constraints. To ensure computational efficiency, the algorithm employs an \textit{early exit mechanism}. During a single CoProposer inference, the LLM evaluates up to $Att\_max$ candidate proposals through rollout-simulated verification. It retains all feasible candidates together with their estimated cumulative rollout rewards. After the bounded search, the feasible candidate with the highest $R(\tau_n)$ is selected. If no candidate is fully feasible, the highest-scoring candidate satisfying the immediate safety constraints is returned.

The computational complexity of the CoProposer is dominated by the number of LLM inference calls. Let $Att_{\text{max}}$ be the maximum number of validation attempts and $K$ be the rollout horizon. The bounded internal rollout reasoning requires at most $O(Att_max * K)$ simulated reasoning steps, while the entire candidate search is elicited within a single CoProposer inference call.
To mitigate latency in real-world applications, we set $Att_{\text{max}}=10$ for all tasks. The replacement of estimated scores beyond $t+1$ with binary feasibility
indicators further reduces the overhead of reward calculation.

\section{Scope of Theoretical Assumptions}
\label{app:scope}

Eq.~\ref{eq:thm1} analyzes the mean-field approximation error under the domain's true reward function $R_n$, i.e., the error introduced by substituting the true neighborhood states $s^t_{\mathcal{N}_n}$ with the mean-field estimate $\mu_n$. We clarify the scope and assumptions below.

\textbf{Domain Motivation.}
The continuous physical dynamics underlying Cooperative Platoon Planning (CPP) and Pandemic Control (PC) motivate the use of a smooth true-return model or a continuous surrogate. However, the differentiability of the physical dynamics alone does not guarantee that the induced return function is twice continuously differentiable. We therefore impose the smoothness and bounded-Hessian conditions below as analytical assumptions on the domain true return function $\mathcal{R}_n$. These assumptions do not apply to the LLM-estimated rollout score $\widehat{R}_n$.

\paragraph{Formal Assumptions.}
For the theoretical analysis, we normalize the positive weights without loss of generality such that
$\sum_{m\in\mathcal{N}_n} w_{n,m}=1,\qquad
w_{\min}:=\min_{m\in\mathcal{N}_n}w_{n,m}>0.$
We assume that $\mathcal{R}_n$ is twice continuously differentiable and that
$
\left\|\nabla^2\mathcal{R}_n(\mathbf{z})\right\|_2
\leq L_2
$
on the relevant line segment. For Eq.~\ref{eq:thm1}, we additionally assume that the first-order Taylor term vanishes at the global mean-field point:
$
\nabla\mathcal{R}_n(\boldsymbol{\mu}_n)^\top
\left(
\mathbf{s}_{\mathcal{N}_n}^{t}
-
\boldsymbol{\mu}_n
\right)=0.
$

\paragraph{Scope Clarification.}
Eq.~\ref{eq:thm1} characterizes the theoretical relationship between neighborhood variance $\sigma^2_n$ and the mean-field approximation quality under the domain's true reward function. This analysis establishes $\sigma^2_n$ as a principled uncertainty indicator, motivating the framework's design choice of transmitting $\sigma^2_n$ during negotiation. In practice, the LLM's rollout-based reward estimation introduces a separate source of approximation that is outside the scope of Eq.~\ref{eq:thm1}. This estimation gap is mitigated through progressive constraint relaxation in rollout verification (Sec.~\ref{sec:macro-1}), which simplifies reward signals beyond $t+1$ to binary feasibility indicators, reducing sensitivity to prediction imprecision.

\section{Proof of Theorem~3.1}
\label{app:tighter_bound}
\begin{proof}
    Let the neighborhood $\mathcal{N}_n$ be partitioned into $P$ disjoint subgroups $\{\mathcal{N}_n^{(p)}\}_{p=1}^P$.
    For each subgroup $p$, define the local weighted mean $\mu_n^{(p)} = \sum_{m \in \mathcal{N}_n^{(p)}} w_{n,m}^{(p)}\, s_m^t$ and local weighted variance $\sigma_n^{2,(p)} = \sum_{m \in \mathcal{N}_n^{(p)}} w_{n,m}^{(p)}\, \|s_m^t - \mu_n^{(p)}\|_2^2$, where $w_{n,m}^{(p)} = w_{n,m} / W^{(p)}$ are the renormalized weights within subgroup $p$ and $W^{(p)} = \sum_{m \in \mathcal{N}_n^{(p)}} w_{n,m}$. 
    By Taylor's theorem applied to the full neighborhood-state
vector around the topology-aware mean-field point, we have
\begin{equation}
\begin{split}
&
\mathcal{R}_n(\mathbf{s}_{\mathcal{N}_n}^{t})
-
\mathcal{R}_n(\boldsymbol{\mu}_n^{(1:P)})
\\
&=
\nabla\mathcal{R}_n
\bigl(\boldsymbol{\mu}_n^{(1:P)}\bigr)^\top
\left(
\mathbf{s}_{\mathcal{N}_n}^{t}
-
\boldsymbol{\mu}_n^{(1:P)}
\right)+
\\
&\frac{1}{2}
\left(
\mathbf{s}_{\mathcal{N}_n}^{t}
-
\boldsymbol{\mu}_n^{(1:P)}
\right)^\top
\nabla^2\mathcal{R}_n(\boldsymbol{\xi}_n)
\left(
\mathbf{s}_{\mathcal{N}_n}^{t}
-
\boldsymbol{\mu}_n^{(1:P)}
\right),
\end{split}
\end{equation}
where $\boldsymbol{\xi}_n$ lies on the line segment between
$\mathbf{s}_{\mathcal{N}_n}^{t}$ and
$\boldsymbol{\mu}_n^{(1:P)}$.

By the first-order cancellation condition in
Theorem~\ref{thm:tighter_bound} and the $L_2$-smoothness
assumption, we obtain
\begin{equation}\label{equ:thm2}
\begin{split}
&
\left|
\mathcal{R}_n(\mathbf{s}_{\mathcal{N}_n}^{t})
-
\mathcal{R}_n(\boldsymbol{\mu}_n^{(1:P)})
\right|
\\
&\leq
\frac{L_2}{2}
\sum_{p=1}^{P}
\sum_{m\in\mathcal{N}_n^{(p)}}
\|s_m^t-\mu_n^{(p)}\|_2^2
\\
&\leq
\frac{L_2}{2w_{\min}}
\sum_{p=1}^{P}
\sum_{m\in\mathcal{N}_n^{(p)}}
w_{n,m}\|s_m^t-\mu_n^{(p)}\|_2^2.
\end{split}
\end{equation}

    By the law of total variance, for the globally normalized weights
$\sum_{m\in\mathcal{N}_n} w_{n,m}=1$, we have
\begin{equation}
\begin{split}
\sigma_n^2
&=
\sum_{m \in \mathcal{N}_n} w_{n,m}\|s_m^t-\mu_n\|_2^2 \\
&=
\underbrace{\sum_{p=1}^{P} W^{(p)}\sigma_n^{2,(p)}}_{\text{within-subgroup variance}}
+
\underbrace{\sum_{p=1}^{P} W^{(p)}\|\mu_n^{(p)}-\mu_n\|_2^2}_{\text{between-subgroup variance}\geq 0}.
\end{split}
\end{equation}
Therefore,
\begin{equation}
\sum_{p=1}^{P} W^{(p)}\sigma_n^{2,(p)}
\leq \sigma_n^2 .
\end{equation}

Furthermore,
\begin{equation}
\sum_{m \in \mathcal{N}_n^{(p)}} w_{n,m}\|s_m^t-\mu_n^{(p)}\|_2^2
= W^{(p)}\sigma_n^{2,(p)} .
\end{equation}
Substituting this into Eq.~\ref{equ:thm2} gives
\begin{equation}
\begin{split}
&\left| \mathcal{R}_n(\mathbf{s}_{\mathcal{N}_n}^{t})
- \mathcal{R}_n(\boldsymbol{\mu}_n^{(1:P)}) \right| \\
&\leq
\frac{L_2}{2\,w_{\min}}
\sum_{p=1}^{P} W^{(p)}\sigma_n^{2,(p)}
\leq
\frac{L_2}{2\,w_{\min}}\sigma_n^2 .
\end{split}
\end{equation}
The second inequality is strict when the between-subgroup variance is positive, i.e., when subgroup means differ under positive subgroup weights. Thus, under the stated first-order cancellation condition, topology-aware partitioning removes the between-subgroup variance from the variance-based upper bound, yielding a no-larger bound, which is strictly smaller when the between-subgroup variance is positive.
\end{proof}

\section{Proof of Proposition~3.2}\label{app:nego}
\begin{proof}
    $\mathcal{G} = (\mathcal{V}, \mathcal{E})$ represents a connected communication graph, and  $r$ denotes the maximum number of negotiation rounds.

\paragraph{Bounded Termination.} At each round $\ell$, the negotiation loop evaluates two termination conditions:
\begin{itemize}
    \item \textit{Early stopping via local consensus:} If $\|P_n^{(\ell)} - P_m^{(\ell)}\| < \delta$ for all  $m \in \mathcal{N}_n$, where $\delta$ is a predefined task-specific threshold (e.g., $\delta = 0.1\,\text{m/s}^2$ for CPP), the agent terminates negotiation.
    \item \textit{Maximum round limit:} The loop terminates after at most $r$ rounds regardless of consensus status.
\end{itemize}
Hence, regardless of whether local consensus is reached, the maximum round limit ensures that the negotiation terminates within $r$ rounds.

\paragraph{Monotonic Non-Decrease of Proposal Quality.}
At the beginning of each negotiation round $\ell + 1$, the best proposal from the previous round, $P_n^{(\ell)}$, is retained in the candidate set for the current round. That is, the candidate set at round $\ell + 1$ is:
\begin{equation}
    \mathcal{C}_n^{(\ell+1)} = \{P_n^{(\ell)}\} 
    \cup \mathcal{C}_n^{\text{new}},
\end{equation}
where $\mathcal{C}_n^{\text{new}}$ denotes the set of newly generated proposals incorporating neighbor information received at round $\ell + 1$.

Since $P_n^{(\ell)} \in \mathcal{C}_n^{(\ell+1)}$ by the Design Invariant, and agent $n$ selects $P_n^{(\ell+1)} = \arg\max_{P \in \mathcal{C}_n^{(\ell+1)}} \mathcal{R}(\tau_n(P))$, we have $\mathcal{R}(\tau_n^{(\ell+1)}) = \max_{P \in \mathcal{C}_n^{(\ell+1)}} \mathcal{R}(\tau_n(P)) \geq \mathcal{R}(\tau_n(P_n^{(\ell)})) =\mathcal{R}(\tau_n^{(\ell)}).$

Thus, the sequence $\{\mathcal{R}(\tau_n^{(\ell)})\}_{\ell=0}^{r}$ is non-decreasing. Under the idealized rational-inference assumption, this score ordering is consistent with rollout quality, so proposal quality is also non-decreasing across negotiation rounds.
The cumulative rollout reward is upper-bounded by the maximum achievable return over the rollout horizon $k$:
\begin{equation}
    \mathcal{R}(\tau_n^{(\ell)}) 
    \leq \sum_{\iota=0}^{k} \gamma^\iota R_{\max} 
    = R_{\max} \cdot \frac{1 - \gamma^{k+1}}{1 - \gamma},
\end{equation}
where $R_{\max} = \sup_{o_n, a_n} |R_n(o_n, a_n)|$ is the single-step reward bound. Therefore, the rollout score remains bounded and non-decreasing over the $r$ negotiation rounds.
\end{proof}

\section{Equations for Welford-based State Aggregation}
The update rules for the weighted mean and variance are defined as follows:
\label{app:welford}
\begin{equation}\label{equ:mean}
\mu_n' = \mu_m + \frac{w_n}{W_n}(s_n^t - \mu_m), 
\end{equation} 
\begin{equation}\label{equ:var}
{\sigma_n^2}' = \frac{W_m}{W_n}\sigma_m^2 + \frac{w_n}{W_n}(s_n^t - \mu_m)(s_n^t - \mu_n'), 
\end{equation}
where $W_n = W_m + w_n$ represents the updated weight. The aggregated mean $\mu_n'$ and variance ${\sigma_n^2}'$ are then abstracted into a textual description of a virtual node within the query prompt, enabling the LLM to analyze trends in the unobserved states. While equations (\ref{equ:mean}) and (\ref{equ:var}) illustrate the atomic update, in practice, agents apply this iteratively or employ parallel variance merger algorithms~\cite{chan1983algorithms} when aggregating disjoint sets.

\section{Simulation Environments and Task Configurations}
\label{app:env_detail}

Table~\ref{tab:hyperparams} summarizes the key settings of the MACRO-LLM framework used across all experiments.
\begin{table}[h]
\centering
\small
\caption{Framework hyperparameters.}
\label{tab:hyperparams}
\begin{tabular}{lc}
\toprule
Parameter & Value \\
\midrule
Rollout horizon $k$ & 2 \\
Max negotiation rounds $r$ & 3 \\
Max proposal attempts $\text{Att}_{\max}$ & 10 \\
Mean-field weight $w_{n,m}$ & Inverse hop distance \\
Consensus threshold $\epsilon$ & 0.02 (CPP) / 0 (PC) \\
Temperature / top-p & 0.3 / 1.0 \\
\bottomrule
\end{tabular}
\end{table}

In MACRO-LLM, the rollout horizon $k=2$ balances verification depth against prompt length: the first rollout step performs precise constraint verification on the immediate next state, while the second step applies relaxed feasibility checks to ensure short-term safety (Sec.~\ref{sec:macro-2}). The negotiation rounds $r=3$ are set based on network diameter $D$: since each agent begins with 1-hop neighbor information and each round propagates aggregated statistics by one additional hop, $r$ rounds yield effective coverage of $r+1$ hops. For the PC topologies ($D \leq 4$), $r=3$ provides full network coverage. For the CPP linear topology ($D=7$), the topology-aware partitioning into predecessor and follower subgroups reduces the effective propagation distance to $D/2$ per direction, which $r=3$ sufficiently covers. 

We evaluate our framework on two complementary domains, CPP and PC, to assess its collaborative performance and architecture-level transfer across the two evaluated simulator domains, utilizing high-fidelity simulators for each task. The evaluation covers continuous and discrete actions, linear and graph topologies, fine-grained and long-horizon planning.

\subsection{Cooperative Platoon Planning (CPP)}
The CPP task uses a vehicle-dynamics environment adapted from the CACC task implementation in Networked-MB-MARL~\cite{ma2024efficient}\footnote{
\href{https://github.com/CDM1619/Networked-MB-MARL}
{Networked-MB-MARL repository}.
}.
\begin{itemize}
    \item \textbf{Topology:} We model a platoon of eight controlled vehicles ($N=8$) on a single-lane highway segment. Each simulation lasts 60 seconds and consists of 120 control steps ($\Delta t=0.5$ s). Each agent operates with one-hop local perception, observing its own state and the states (e.g., headway and velocity) of its immediate neighboring vehicles. Due to the linear topology, information from preceding and following vehicles is aggregated separately using direction-specific statistics (Theorem~\ref{thm:tighter_bound}).
    \item \textbf{Objective:} The benchmark objective is for agents to produce coordinated high-level acceleration directives that stabilize the platoon-level state toward an optimal headway of 20 m and a target velocity of 15 m/s.
    \item \textbf{Scenarios:} We evaluate performance in two dynamic settings: 
    (1) \textit{Scenario 1: Catch-up}: The leading vehicle accelerates to close the gap with a target vehicle, testing whether the coordination framework can recover stable formation under a large velocity perturbation.
    (2) \textit{Scenario 2: Slow-down}: The platoon begins at an elevated velocity, while the leading reference vehicle decelerates, challenging the agents to maintain coordinated and stable state transitions under a deceleration perturbation.
\end{itemize}

\subsection{Pandemic Coordination (PC)}

\begin{figure}[htbp]
\centering
\subfloat[]{\includegraphics[width=0.9\linewidth]{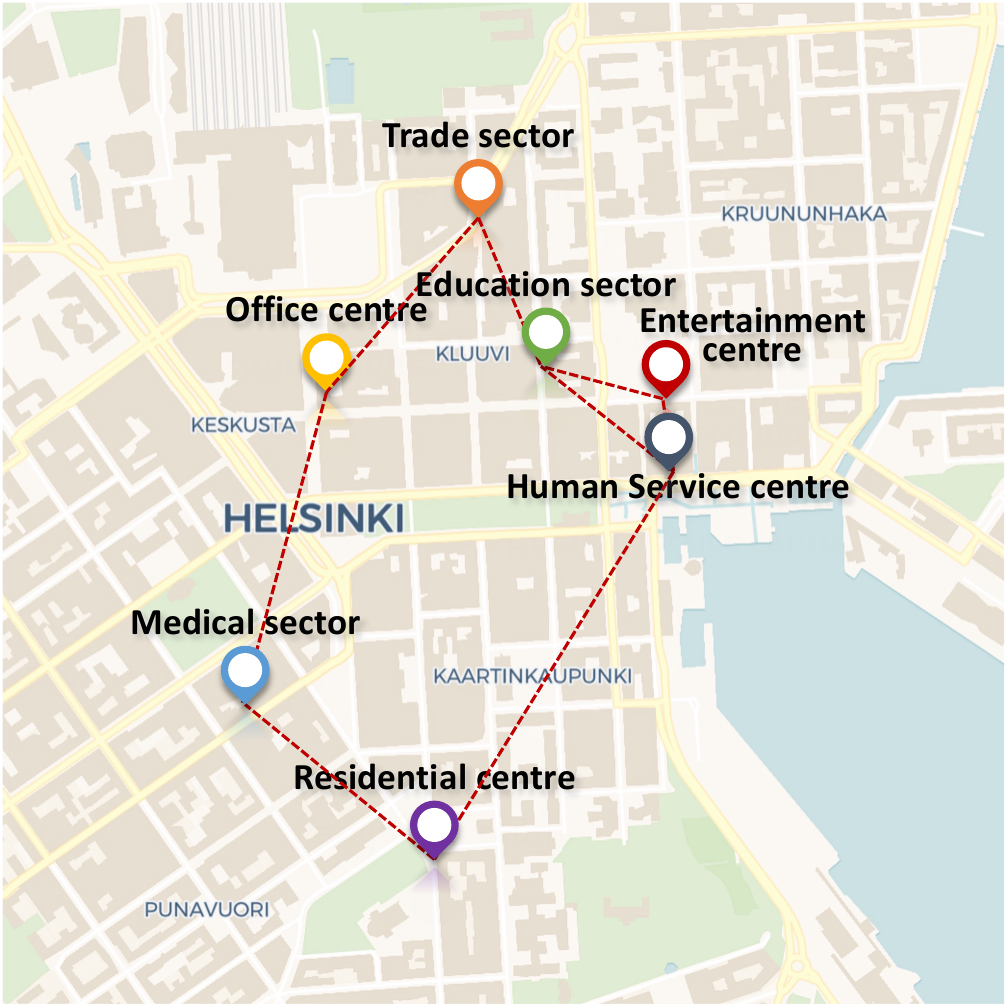}\label{exp:pan-map-hel}}
\hfil
\subfloat[]{\includegraphics[width=0.9\linewidth]{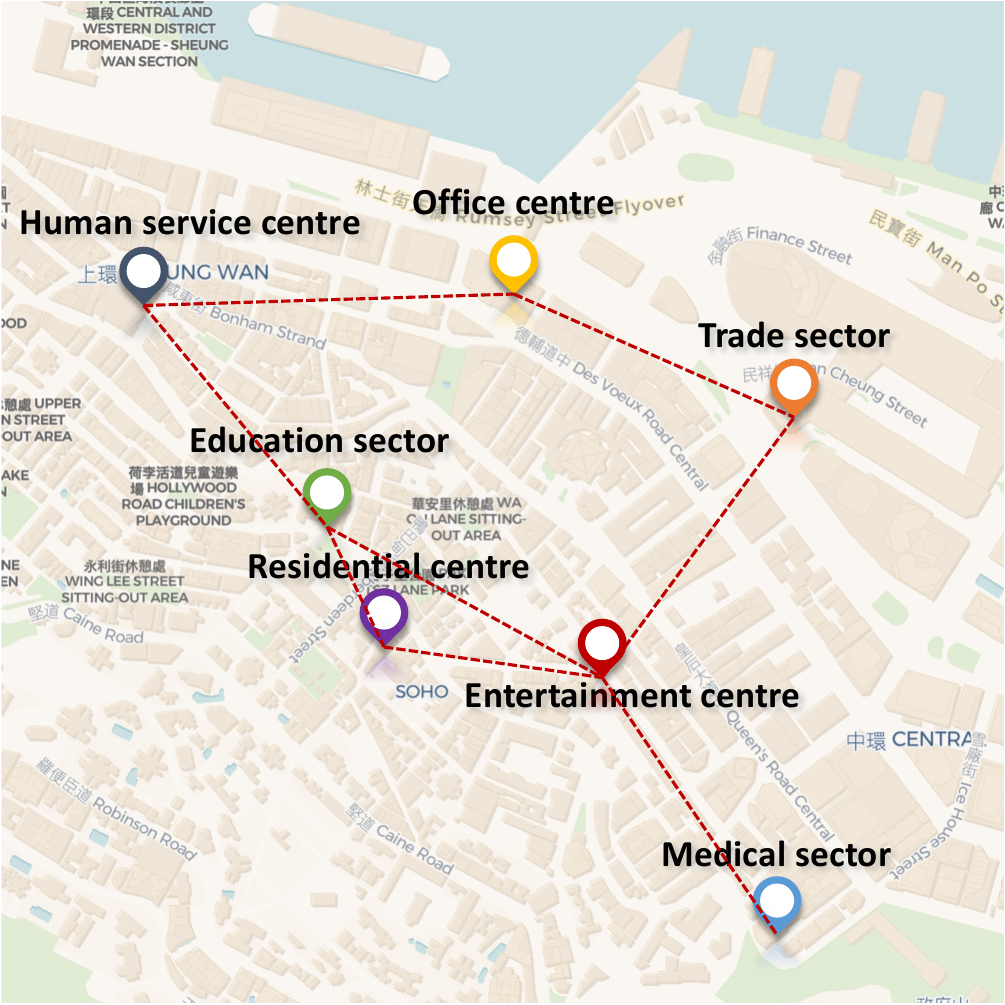}\label{exp:pan-map-hk}}
\hfil
\subfloat[]{\includegraphics[width=0.9\linewidth]{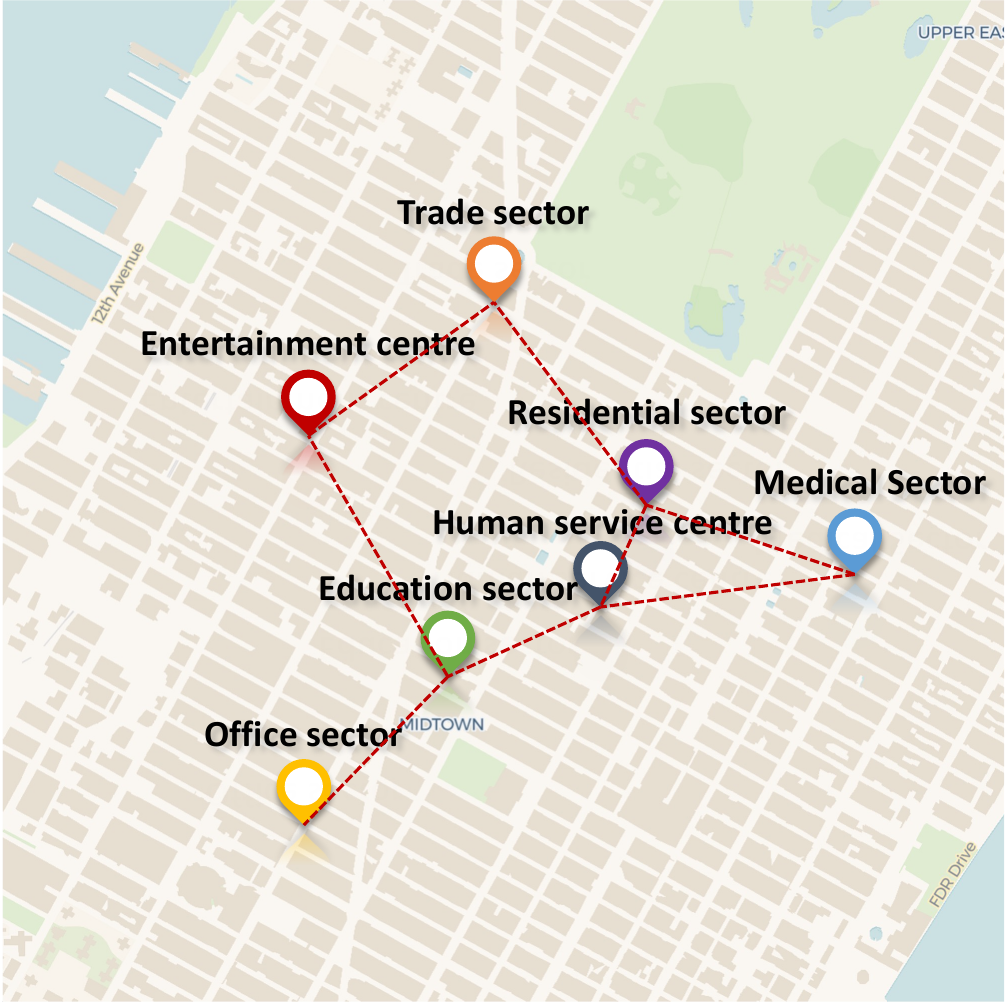}\label{exp:pan-map-ny}}
\caption{Communication topology of cities for pandemic networks. (a) Helsinki (b) Hong Kong (c) New York.}
\label{fig:EC-map}
\end{figure}

The PC task is simulated using the open-source Pandemic Simulator\footnote{
\href{https://github.com/SonyResearch/PandemicSimulator}
{Pandemic Simulator repository},
under Apache-2.0.
}~\cite{kompella2020reinforcement}.
\begin{itemize}
\item \textbf{Topology:} We model the urban environment as a graph of seven agents representing distinct facility types: {Residential sector (Home), Office sector (Office), Education sector (School), Medical sector (Hospital), Trade sector (Retail and Grocery), Entertainment centre (Restaurant, Bar and Hair), and Human service centre (Government)}. 
To construct realistic urban topologies, we utilize Google Maps to identify clusters of functionally similar buildings within specific urban blocks, marking their centroids as individual agents. Communication links are established based on spatial proximity, forming a connected topology as shown in Fig.~\ref{fig:EC-map}. Since the urban graph does not admit a natural disjoint partition of each agent's neighborhood, mean-field statistics are computed over the full neighborhood without subgroup decomposition. The simulation spans 120 days, during which agents operate under partial observability regarding global infection rates.

\item \textbf{Objective:} The agents dynamically adjust containment policies, such as restrictions on gatherings and business operating hours, to limit viral spread while maintaining essential socioeconomic activity. For the PC task, the consensus threshold is set to $\delta=0$, requiring exact agreement.

\item \textbf{Scenarios:} We evaluate the framework's generalization capabilities across three distinct urban settings with varying population scales:
(1) \textit{Helsinki}: A standard community setting with a population of 500.
(2) \textit{Hong Kong}: A denser urban setting with a population of 1000.
(3) \textit{New York}: A high-density complex topology with a population of 1500.
\end{itemize}

\section{Baseline Methods}\label{app:baselines}
To evaluate MACRO-LLM, we implement and adapt the following baselines for both the CPP and PC tasks.

\subsection{MARL Baselines}
\label{app:marl_baselines}

\paragraph{Baseline selection.} We compare against three MARL methods chosen to span complementary design choices relevant to coordination under partial observability: (1) \textbf{DPPO}~\cite{ma2024efficient}, a decentralized PPO-style policy optimization method; (2) \textbf{DMPO}~\cite{ma2024efficient}, a recent model-based method for large-scale networked control; and (3) \textbf{IC3Net}~\cite{ic3net2018learning}, a canonical communication-aware MARL baseline. DPPO and DMPO are adopted from Ma et al.~\cite{ma2024efficient}; we use their simulators, the original CPP topology, and their released hyperparameters where applicable. All methods on a given task share the same topology, observation and action interfaces, and training protocol.

\paragraph{Architecture and capacity.} DPPO is model-free with MLP[64,64] actor and critic networks. DMPO augments MLP[64,64] policy/value networks with a one-layer GNN world model (residual edge/node embeddings; edge/node hidden sizes [512,512] for CPP and [24,24] for PC). IC3Net uses an MLP policy/value network with a communication module (MLP[64,64] for CPP, MLP[128,128] for PC) and GNN edge/node hidden sizes [16,16]. All methods use discount $\gamma=0.99$ and a PPO clip ratio of $\epsilon=0.2$ (with $\epsilon=0.15$ for IC3Net on CPP), and a target KL of $0.01$ for DPPO and $7.5\times10^{-3}$ for DMPO and IC3Net. For learning rates, DPPO uses 5e-5\,/\,5e-4 for the policy/value networks; DMPO uses a world-model rate of 2e-4, a value rate of 5e-4, and a policy rate of 5e-5 (CPP) or 5e-4 (PC); IC3Net uses policy/value rates of 5e-4\,/\,1e-4 and a communication rate of 1e-2. Hyperparameters follow the released configurations of Ma et al.~\cite{ma2024efficient} where applicable; where training stability required, we adjusted the learning rates and report the final values above. Full configurations are released with our code.

\paragraph{Training budget and cost.} Each baseline is trained for 25{,}000 iterations, corresponding to 15M environment steps on CPP and 3M on PC. Training is computationally substantial: on a single RTX 3090, DPPO requires approximately 39\,h (CPP) and 57.5\,h (PC), and DMPO approximately 75\,h (CPP) and 150\,h (PC). This cost is incurred per task and per topology, since MARL policies require retraining whenever the topology or agent count changes. Each trained baseline is evaluated over 5 episodes per seed across 5 independent runs.

\paragraph{Convergence diagnostics.} We follow standard RL evaluation practice~\cite{henderson2018deep, agarwal2021deep} and assess late-stage stabilization via the reward distribution over the last 10\% of training and the normalized reward slope over the last 20\%. As reported in Table~\ref{tab:marl_convergence}, the near-zero slopes ($|\text{slope}|\leq 5\times10^{-4}$), together with KL divergence remaining within target thresholds throughout training, confirm asymptotic stabilization.
\begin{table}[t]
\centering
\small
\setlength{\tabcolsep}{4pt}
\caption{MARL convergence diagnostics on representative scenarios. Values report the last-10\% reward (mean\,$\pm$\,std); slope denotes the normalized reward trend over the last 20\% of training. KL divergence remained within target thresholds throughout training.}
\label{tab:marl_convergence}
\resizebox{\linewidth}{!}{\begin{tabular}{lccc}
\toprule
\textbf{Scenario} & \textbf{DPPO} & \textbf{DMPO} & \textbf{IC3Net} \\
\midrule
CPP / CU
& \makecell{$-19.32\pm5.97$\\(slope: $2\times10^{-5}$)}
& \makecell{$-72.79\pm14.14$\\(slope: $3\times10^{-4}$)}
& \makecell{$-52.76\pm10.22$\\(slope: $-1\times10^{-6}$)} \\
\midrule
PC / Hel
& \makecell{$-56.78\pm11.11$\\(slope: $-9\times10^{-5}$)}
& \makecell{$-51.92\pm11.43$\\(slope: $2\times10^{-6}$)}
& \makecell{$-51.53\pm12.05$\\(slope: $9\times10^{-6}$)} \\
\bottomrule
\end{tabular}}
\end{table}

\subsection{LLM-based MAS Baselines}
We compare against recent frameworks utilizing LLMs for collaboration: (1) \textbf{ToM-Belief}~\cite{li2023theory}: A MAS that employs second-order Theory of Mind for introspection via belief state updates and environmental feedback; (2) \textbf{ChatEval}~\cite{chan2023chateval}: Adopts the Simultaneous-Talk-with-Summarizer protocol (Algorithm 3), which is structurally similar to our task setting; (3) \textbf{LAMEN}~\cite{davidson2024evaluating}: A framework that constructs internal ``mental notes'' alongside public messages to facilitate strategic negotiation.

\subsection{Adaptation, Fairness, and System Design}
\label{app:system_design}
To ensure a fair and reproducible comparison, we maintain experimental consistency across all LLM-based methods. In MACRO-LLM, the LLM performs all strategic reasoning: generating collaborative proposals, predicting future observations during rollout verification, evaluating proposal conflicts and updating spatial strategies in the Negotiator, and diagnosing performance drops in the Introspector. Deterministic computations independent of the LLM reasoning process, such as mean-field statistics (Eqs.~\ref{eq:mean_field_observation}--\ref{eq:mean_field_variance}) and drift intensity (Eq.~\ref{eq:drift_intensity}), are offloaded to an external numerical module to ensure precision. All inference-dependent computations, including state prediction and constraint evaluation during rollout verification, are performed entirely by the LLM. LLM outputs are validated against a structured XML schema; outputs failing validation trigger an automatic retry (up to 3 attempts).

All LLM-based baselines (ToM-Belief, ChatEval, LAMEN) share the same backbone model (GPT-4o), environmental interfaces, external numerical module, schema validation, and retry mechanism as MACRO-LLM. Methods differ only in their prompting strategies and inter-agent communication protocols (e.g., mental notes generation in LAMEN, belief updates in ToM-Belief). This ensures that performance differences reflect methodological contributions rather than implementation advantages. MARL and LLM-based methods follow different learning and inference regimes: MARL baselines are task-specialized policies trained on each topology, whereas LLM-based methods coordinate zero-shot from pre-trained knowledge. We therefore treat the MARL baselines not as compute-normalized competitors but as task-specialized reference points. The comparison remains informative because all methods are evaluated under identical simulators, task objectives, partial-observability constraints, and domain metrics (e.g., SD-H, infection rate). Under this protocol, matching or surpassing task-trained MARL references in a zero-shot setting is practically meaningful for dynamic deployments where per-scenario retraining is costly.


\subsection{Classical Controller References}
\label{app:controller_reference}

We further compare MACRO-LLM with four task-specific platooning controllers on CPP Catch-up: proportional--derivative control (PD)~\citep{swaroop1999constant}, proportional--integral--derivative control (PID)~\citep{gunagwera2022longitudinal}, the intelligent driver model (IDM)~\citep{treiber2000congested}, and cooperative adaptive cruise control (CACC)~\citep{milanes2014modeling}. All methods are evaluated in the same CPP Catch-up environment using the same metrics. The classical controllers use domain-specific control laws with parameters tuned for this task. MACRO-LLM uses the CPP task interface and self-checking rules without LLM parameter updates. As shown in Table \ref{tab:classical_cpp}, MACRO-GPT achieves the lowest RMSE-H and SD-H, while IDM and PID obtain lower velocity errors, indicating competitive headway coordination without dominance across all metrics. 

\begin{table}[t]
\centering
\small
\setlength{\tabcolsep}{3pt}
\caption{Comparison with task-specific platooning controllers on CPP Catch-up. Lower values are better. Best and second-best results are shown in bold and underlined, respectively.}
\label{tab:classical_cpp}
\begin{tabular}{lcccc}
\toprule
Method & RMSE-H & RMSE-V & SD-H & SD-V \\
\midrule
PD          & 1.894 & 1.149 & 1.084 & 0.562 \\
PID         & 1.467 & \underline{0.508} & 0.726 & \textbf{0.153} \\
IDM         & \underline{1.305} & \textbf{0.502}
            & \underline{0.449} & \underline{0.166} \\
CACC        & 1.446 & 0.723 & 0.750 & 0.358 \\
MACRO-GPT   & \textbf{1.237} & 0.721
            & \textbf{0.409} & 0.203 \\
MACRO-Llama & 1.412 & 0.779 & 0.645 & 0.365 \\
\bottomrule
\end{tabular}
\end{table}

\section{Case Study: Hierarchical Deployment for Coarse-Grained Strategic Coordination} \label{app:deploy}
As a general-purpose strategic coordination framework, MACRO-LLM outputs high-level collaborative directives rather than low-level control commands. For domains whose environment states evolve faster than deliberative LLM reasoning, MACRO-LLM can be used as a coarse-grained strategic coordination layer above a domain-specific execution layer~\cite{liu2025colmdriver,wen2024diluknowledge}. In this section, we describe the deployment architecture and experimentally validate its feasibility through a decision frequency analysis on the CPP task. 

\begin{figure}[htbp]
    \centering
    \includegraphics[width=1.01\linewidth]{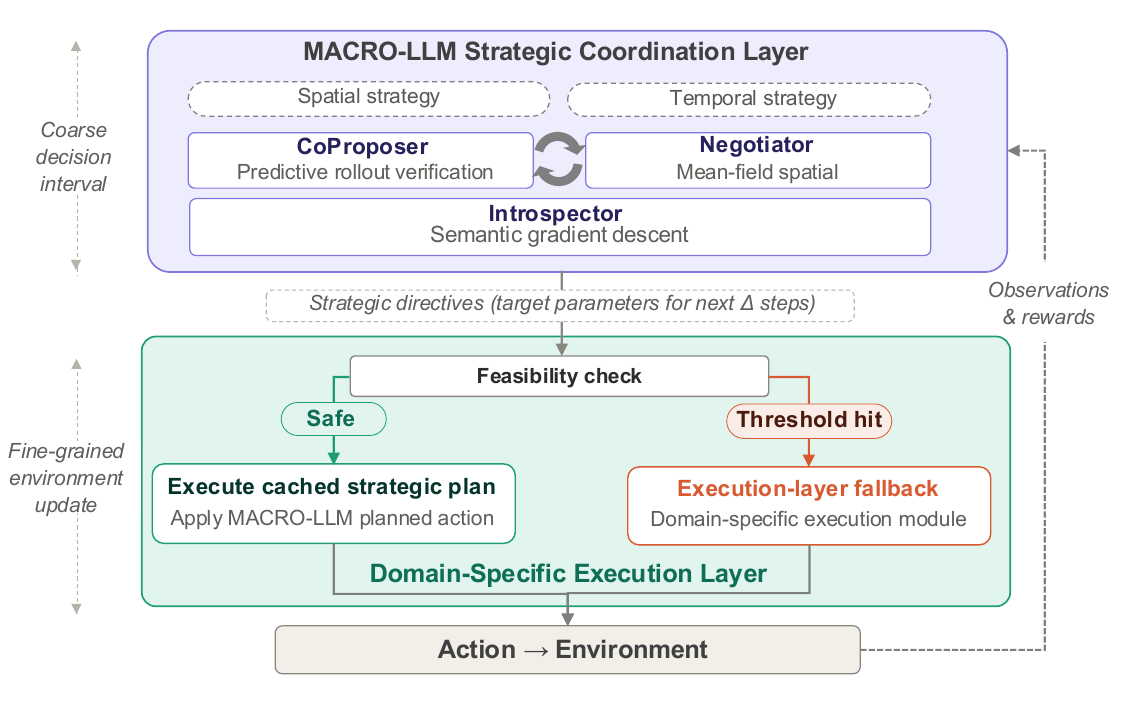}
    \caption{The strategic coordination layer executes a negotiation cycle every $\Delta$ steps to produce collaborative directives, while the domain-specific execution layer updates the environment at finer granularity using cached directives and task-specific feasibility checks.}
    \label{fig:hierarchical}
\end{figure}

\subsection{Hierarchical Deployment Paradigm}
\label{app:hierarchical:paradigm}
As illustrated in Fig.~\ref{fig:hierarchical}, the hierarchical paradigm consists of two layers:
\begin{enumerate}
    \item \textit{Strategic Coordination Layer (Second-level granularity)}. MACRO-LLM operates at a coarser temporal granularity, executing a full negotiation cycle among agents every $\Delta$ time steps. At each decision epoch, it produces strategic directives by reasoning about unobservable states and negotiating with neighboring agents. Between decision epochs, the execution layer carries out the planned actions.
    \item \textit{Domain-Specific Execution Layer (Millisecond-level granularity)}. At each time step, a lightweight controller checks whether the current state satisfies predefined safety thresholds. If so, it follows the cached strategic plan from the latest MACRO-LLM decision. Otherwise, a domain-specific safety controller (e.g., PID for CPP) immediately overrides the planned action to maintain safety.
\end{enumerate}
MACRO-LLM requires no architectural modification to operate within this paradigm---its outputs are inherently strategic directives that can be integrated with a domain-specific execution layer, as demonstrated here in the CPP simulator.
 
\subsection{Decision Frequency Analysis}
\label{app:hierarchical:experiment}
 
To validate the feasibility of hierarchical deployment, we evaluate MACRO-LLM on the CPP Catch-up scenario under varying decision intervals $\Delta \in \{1, 3, 5, 10\}$, where $\Delta=1$ corresponds to the original setting used in our main experiments.
At each decision epoch, MACRO-LLM produces a strategic action spanning the subsequent $\Delta$ time steps. Between decision epochs, the execution layer applies the most recently cached coordination directive and uses a simple task-specific feasibility rule to keep the simulated trajectory within valid benchmark constraints.
Table~\ref{tab:hierarchical} presents the results alongside representative baselines from Table~\ref{tab:cacc} for reference.
 

\begin{table}[h]
\centering
\caption{Decision frequency analysis on the CPP Catch-up scenario. DPPO and LAMEN from Table~\ref{tab:cacc} for reference. All metrics $\downarrow$.}
\label{tab:hierarchical}
\small
\resizebox{\linewidth}{!}{%
\begin{tabular}{lccccc}
\toprule
\textbf{Config.} & \textbf{Red.} & \textbf{RMSE-H} & \textbf{RMSE-V} & \textbf{SD-H} & \textbf{SD-V} \\
\midrule
DPPO & -- & 1.603 & 0.978 & 0.994 & 0.474 \\
LAMEN & -- & 1.891 & 0.648 & 3.320 & 0.601 \\
\midrule
\textbf{Hier.} $\Delta=10$ & 90\% & 1.443 & 0.623 & 0.634 & 0.225 \\
\textbf{Hier.} $\Delta=5$  & 80\% & 1.319 & 0.721 & 0.524 & 0.281 \\
\textbf{Hier.} $\Delta=3$  & 67\% & 1.408 & 0.596 & 0.564 & 0.234 \\
\midrule
\textbf{MACRO} $\Delta=1$ & 0\% & 1.237 & 0.721 & 0.409 & 0.203 \\
\bottomrule
\multicolumn{6}{l}{\textit{Hier.}: hierarchical deployment;} \\
\multicolumn{6}{l}{$\Delta$: decision interval (time steps between MACRO-LLM invocations)}\\
\multicolumn{6}{l}{\textit{Red.}: LLM calls reduced relative to $\Delta=1$.}
\end{tabular}}
\end{table}
 
All reduced-frequency configurations remain within the benchmark’s validity constraints in our simulations, confirming that the execution layer's safety override mechanism provides reliable protection between decision epochs.
As shown in Table \ref{tab:hierarchical}, even with 90\% fewer LLM invocations ($\Delta=10$), the hierarchical deployment consistently outperforms both DPPO and LAMEN across all metrics. The SD-H of hierarchical deployment ($\Delta=10$) is 36.2\% lower than that of DPPO and 80.9\% lower than that of LAMEN.
This demonstrates that MACRO-LLM's coordination reasoning provides substantial value even at highly reduced decision frequencies, as the strategic directives continue to guide inter-agent collaboration during the intervals between LLM invocations.
Furthermore, increasing the decision frequency yields overall improvements in coordination metrics. SD-H improves from 0.634 ($\Delta=10$) to 0.409 ($\Delta=1$), a 35.5\% reduction, confirming that the performance gains stem from MACRO-LLM's collaborative reasoning. 
These results indicate that MACRO-LLM can be effectively deployed within a hierarchical paradigm, achieving competitive coordination quality while substantially reducing computational overhead.

\section{Detailed Evaluation Metrics}
\label{app:metrics}
We provide the detailed definitions and mathematical formulations of the metrics used in our experiments.
\subsection{CPP Task Metrics}
Following previous works~\cite{rmse2011cooperative, sd2006impact}, we evaluate the platoon efficiency and stability using:
\begin{itemize}
    \item \textbf{Root Mean Square Error (RMSE-H/V)}: Measures the deviation between the actual headway/velocity and the target values, averaged across all following vehicles ($i=1 \dots N$) over the simulation steps ($t=1 \dots T$). 
    For example, RMSE of Headway (RMSE-H) is calculated as:
    \begin{equation}
        \text{RMSE-H} = \sqrt{\frac{1}{N \cdot T} \sum_{t=1}^{T} \sum_{i=1}^{N} (h_i^t - h_\text{target})^2},
    \end{equation}
    where $h_i^t$ is the headway of vehicle $i$ at time $t$, and $h_\text{target}$ is the optimal safety distance. RMSE-V is computed similarly using velocity deviations as follows:
    \begin{equation}
        \text{RMSE-V} = \sqrt{\frac{1}{N \cdot T} \sum_{t=1}^{T} \sum_{i=1}^{N} (v_i^t - v_\text{leader})^2},
    \end{equation}
    where $v_\text{leader}$ denotes the velocity of the leading vehicle.

    \item \textbf{Standard Deviation (SD-H/V)}: Measures the collaborative consistency of the platoon across all agents. Specifically, at each time step $t$, we calculate the standard deviation of the headway/velocity across all $N$ vehicles. The final reported metric is the temporal average of these spatial standard deviations over the duration $T$:
    \begin{equation}
        \text{SD-H} = \frac{1}{T} \sum_{t=1}^{T} \sqrt{\frac{1}{N} \sum_{i=1}^{N} (h_i^t - \bar{h}^t)^2},
    \end{equation}
    where $\bar{h}^t = \frac{1}{N} \sum_{j=1}^{N} h_j^t$ is the average headway of the entire platoon at time $t$. A lower SD value indicates that all agents are maintaining consistent states simultaneously, reflecting high coordination stability. SD-V is computed similarly using velocity deviations as follows:
    \begin{equation}
        \text{SD-V} = \frac{1}{T} \sum_{t=1}^{T} \sqrt{\frac{1}{N} \sum_{i=1}^{N} (v_i^t - \bar{v}^t)^2}.
    \end{equation}
\end{itemize}

\subsection{PC Task Metrics}
Following Kompella et al.~\cite{kompella2020reinforcement}, we employ the following metrics to assess containment efficacy. Let $P_{total}$ be the total population, and $I^t, C^t, D^t$ be the count of infected, critical, and dead cases on day $t$, respectively.

\begin{itemize}
    \item \textbf{Normalized Infection (I\_n)}: The ratio of the cumulative number of infected cases to the total population: 
    \begin{equation}
        I\_n = \frac{\sum_{t} \Delta I_{new}^t}{P_{total}}.
    \end{equation}
    
    \item \textbf{Normalized Peak Infection (PI\_n)}: Measures the highest pressure on the healthcare system: 
    \begin{equation}
        PI\_n = \frac{\max_{t} (I^t)}{P_{total}}.
    \end{equation}

    \item \textbf{Normalized Dead (D\_n)}: The ratio of total dead cases to the total population at the end of the episode ($t_{end}$):
    \begin{equation}
        D\_n = \frac{D^{t_{end}}}{P_{total}}.
    \end{equation}

    \item \textbf{Pandemic Duration (PD)}: The time span (in days) from the first confirmed infection ($t_{start}$) until the pandemic is effectively contained:
    \begin{equation}
        PD = t_{end} - t_{start}, 
    \end{equation}
    where $t_{end} = \min \{t \mid I^t + C^t = 0, \forall \tau > t\}$.
\end{itemize}

\section{Analysis of Backbone Robustness}\label{app:gen}
\begin{table}[!t]
\centering
\renewcommand{\arraystretch}{1}
\caption{Backbone robustness of MACRO-LLM across six LLM backbones in the two evaluated simulator domains S1/S2: Catch-up/Slow-down scenarios of CPP tasks. C1-C3: PC topologies for Helsinki, Hong Kong, and New York.}
\label{tab:models}
\setlength{\tabcolsep}{6pt}
\resizebox{0.45\textwidth}{!}{\begin{tabular}{
  l l l llll
}
\toprule
\multicolumn{3}{c}{} & \multicolumn{4}{c}{\textbf{Metrics} ($\downarrow$)} \\
\cmidrule(l){4-7}
\multicolumn{2}{c}{\textbf{Tasks}} &  \multicolumn{1}{c}{\textbf{Base Models}}
& \multicolumn{1}{c}{RMSE-H} & \multicolumn{1}{c}{RMSE-V}
& \multicolumn{1}{c}{SD-H} & \multicolumn{1}{c}{SD-V} \\
\midrule
\multirow{12}{*}{\rotatebox{90}{\textbf{CPP}}} & \multirow{6}{*}{\textit{S1}}
& DeepSeek-V3.1 & 1.528 & {0.553} & 0.611 & {0.190} \\
& & Llama-3.3-70B & 1.412 & 0.779 & 0.645 & 0.365 \\
& & Mistral-3.2-24B & 1.891 & 0.784 & 0.997 & 0.326\\
& & Qwen3-flash & 3.521 & 1.910 & 2.935 & 0.579 \\
& & GPT-4o-mini & 2.906 & 1.667 & 1.858 & 0.936 \\
& & GPT-4o & 1.237 & 0.721 & 0.409 & 0.203 \\
\cmidrule(lr){2-7}
& \multirow{6}{*}{\textit{S2}}
& DeepSeek-V3.1 & 0.649 & 3.388 & 0.274 & 0.417\\
& & Llama-3.3-70B & 1.163 & 3.736 & 0.889 & 1.224 \\
& & Mistral-3.2-24B & 1.842 & 3.523 & 1.157 & 0.916\\
& & Qwen3-flash & 3.871 & 3.678 & 2.946 & 0.974 \\
& & GPT-4o-mini & 2.352 & 3.999 & 1.598 & 1.171 \\
& & GPT-4o & 0.496 & 3.310 & 0.128 & 0.238 \\
\addlinespace[2pt]
\midrule
\multicolumn{2}{c}{\textbf{Tasks}} & \multicolumn{1}{c}{\textbf{Base Models}}
& \multicolumn{1}{c}{I\_n} & \multicolumn{1}{c}{PI\_n}
& \multicolumn{1}{c}{D\_n} & \multicolumn{1}{c}{PD} \\
\midrule
\multirow{18}{*}{\rotatebox{90}{\textbf{PC}}} & \multirow{6}{*}{\textit{C1}}
& DeepSeek-V3.1 & 0.098 & 0.038 & 0.002 & 46\\
& & Llama-3.3-70B & 0.013 & 0.008 & 0.000 & 14 \\
& & Mistral-3.2-24B & 0.043 & 0.014 & 0.003 & 31\\
& & Qwen3-flash & {0.010} & {0.008} & 0.002 & 16 \\
& & GPT-4o-mini & {0.010} & 0.010 & {0.000}     & 17 \\
& & GPT-4o & 0.011 & 0.008 & 0.000 & 10 \\
\cmidrule(lr){2-7}
& \multirow{6}{*}{\textit{C2}}
& DeepSeek-V3.1 & 0.008 & 0.005 & 0.001 & 22 \\
& & Llama-3.3-70B & 0.012 & 0.005 & 0.000 & 16 \\
& & Mistral-3.2-24B & 0.009 & 0.004 & 0.001 & 22\\
& & Qwen3-flash & 0.008 & 0.006 & 0.001 & {18} \\
& & GPT-4o-mini & {0.006} & {0.004} & {0.000}     & 24 \\
& & GPT-4o & 0.008 & 0.004 & 0.001 & 23 \\
\cmidrule(lr){2-7}
& \multirow{6}{*}{\textit{C3}}
& DeepSeek-V3.1 & {0.003} & {0.003} & {0.000} & {11}\\
& & Llama-3.3-70B & 0.013 & 0.006 & 0.000 & 25 \\
& & Mistral-3.2-24B & 0.009 & 0.006 & 0.000 & 14\\
& & Qwen3-flash & {0.003} & {0.003} & {0.000} & 13 \\
& & GPT-4o-mini & {0.003} & {0.003} & {0.000}     & 13 \\
& & GPT-4o & 0.004 & 0.003 & 0.000 & 25 \\
\bottomrule
\end{tabular}}
\end{table}

We investigate the robustness of MACRO-LLM across six LLM backbones, including DeepSeek-V3.1~\cite{deepseek}, Llama-3.3-70B-Instruct~\cite{grattafiori2024llama}, Mistral-Small-3.2-24B-Instruct, Qwen3-Flash~\cite{bai2023qwen}, GPT-4o-mini, and GPT-4o. Table~\ref{tab:models} reports the results. In the CPP task, performance differences across backbones are more pronounced, which may reflect the precision required for continuous-valued decisions. In the Catch-up scenario, DeepSeek-V3.1 and GPT-4o reduce RMSE-H by 47.4\% and 57.4\%, respectively, relative to GPT-4o-mini. By contrast, Qwen3-Flash and GPT-4o-mini remain competitive with GPT-4o across the three PC topologies, suggesting that PC is less sensitive to numerical precision. Overall, these results suggest that backbone sensitivity depends on the requirements of each task and that MACRO-LLM can support effective coordination with different LLM backbones.

\section{Computational and Communication Scalability Analysis}
\label{app:scala}

We provide a theoretical and experimental analysis of MACRO-LLM, focusing on communication complexity and computational workload.

\subsection{Communication Complexity Analysis}
Following the definition by Yao~\cite{yao1979some}, communication complexity is quantified by the total number of bits exchanged during the protocol. In our decentralized framework, we focus on the per-agent communication load. Let $\mathcal{G}=(\mathcal{V}, \mathcal{E})$ be the agent communication graph. We denote the average node degree as $\bar{d}$ and the maximum negotiation rounds as $r$. The topology of the CPP task is linear ($\bar{d} \approx 2$). In the PC task, the topology represents an urban graph with average degree $\bar{d}_\text{PC}$. We assume sparse topologies where $\bar{d}$ does not grow linearly with the total number of agents $N$ (i.e., $\bar{d} \ll N$). As the network scales up, MACRO-LLM employs mean-field approximation to compress the states of unobservable agents into a single virtual node representation. Neglecting inference stochasticity, the per-agent message size is theoretically invariant to the network scale, which is represented as $M$. Consequently, the per-agent communication complexity can be represented as $O(\bar{d} \cdot r \cdot M)$.

Since $\bar{d}$, $r$, and $M$ are independent of network scale $N$, the per-agent load remains constant as the network scales in theory. In other words, the per-agent communication load does not increase with the number of agents as long as the topology's connectivity degree remains constant.

Table~\ref{tab:comm_overhead} empirically confirms this analysis and quantifies the benefit of mean-field aggregation. As the platoon scales from $N=8$ to $N=32$, the mean-field per-agent payload stays essentially flat (0.034--0.037M tokens per step), consistent with the predicted scale-invariance, whereas raw-state exchange grows from 0.041M to 0.072M tokens as each agent must encode an increasing number of neighbor states. The resulting savings therefore widen with scale, from 18.8\% at $N=8$ to 52.0\% at $N=32$. This confirms that compressing unobservable agents into a virtual mean-field node keeps per-agent communication bounded as the network grows.

\begin{table}[t]
\centering
\small
\setlength{\tabcolsep}{3pt}
\caption{Communication overhead of mean-field aggregation versus raw-state exchange on the CPP task. Token counts are reported in millions. $\Delta\$$ denotes dollar-cost savings. Savings are computed using GPT-4o pricing and unrounded token counts; displayed token values are rounded to three decimals.}
\label{tab:comm_overhead}
\resizebox{\linewidth}{!}{%
\begin{tabular}{lccccc}
\toprule
\textbf{$N$} & 
\textbf{MF Tok.} & 
\textbf{Raw Tok.} & 
\textbf{$\Delta$ \$/A/St.} & 
\textbf{$\Delta$ \$/A/Ep.} & 
\textbf{Red.} \\
 & 
\textbf{(M/A/St.)} & 
\textbf{(M/A/St.)} & 
 & 
 & 
 \\
\midrule
8  & 0.034 & 0.041 & \$0.019 & \$2.325  & 18.8\%$\downarrow$ \\
16 & 0.037 & 0.054 & \$0.043 & \$5.219  & 32.0\%$\downarrow$ \\
24 & 0.037 & 0.063 & \$0.065 & \$7.796  & 41.5\%$\downarrow$ \\
32 & 0.035 & 0.072 & \$0.093 & \$11.217 & 52.0\%$\downarrow$ \\
\bottomrule
\end{tabular}%
}
\end{table}


\subsection{Computational Efficiency and Latency}
Given that the input prompt length remains bounded as analyzed above, the upper bound of the inference workload for a single agent does not increase with the number of agents. Under API-based deployment, per-step latency is dominated by remote-API network overhead (~5.99 s per GPT-4o call), rather than by the framework’s reasoning complexity. This cost is not intrinsic to the coordination formulation and can be reduced by using lower invocation frequencies or locally deployed lightweight backbones.

\textbf{(1) Local deployment of lightweight models.} 
Table~\ref{tab:models} demonstrates that lightweight models such as GPT-4o-mini and Qwen3-flash achieve competitive coordination performance across both tasks. Table~\ref{tab:cost_backbone} quantifies the corresponding cost: while token consumption per agent stays within a comparable range across backbones, the monetary cost differs by over an order of magnitude. On CPP\,/\,Catch-up, the per-episode cost per agent drops from \$17.18 with GPT-4o to \$1.06 with GPT-4o-mini and \$0.44 with Qwen3-flash, a 16$\times$
and 39$\times$ reduction respectively; a similar 16$\times$ gap holds on PC\,/\,Helsinki (\$13.05 versus \$0.81). Combined with the competitive performance in Table~\ref{tab:models}, this indicates that the coordination quality of MACRO-LLM does not depend on the most expensive backbone. Furthermore, locally deploying these lightweight models eliminates API network latency entirely, reducing per-call inference time by an order of magnitude compared to remote API access.
\begin{table}[t]
\centering
\small
\setlength{\tabcolsep}{4pt}
\caption{Cost across LLM backbones on the CPP and PC tasks. Token counts are reported in millions, as input\,/\,output. A: per agent; St.: per step; Ep.: per episode.}
\resizebox{\linewidth}{!}{\begin{tabular}{lcccc}
\toprule
\textbf{Backbone} & \textbf{Tok./A/St.} & \textbf{Tok./A/Ep.} & \textbf{\$/A/St.} & \textbf{\$/A/Ep.} \\
\midrule
\multicolumn{5}{l}{\textit{(a) CPP / Catch-up ($N=8$ agents)}} \\
\midrule
GPT-4o        & 0.034 / 0.006 & 4.024 / 0.712 & \$0.143 & \$17.181 \\
GPT-4o-mini   & 0.035 / 0.006 & 4.186 / 0.722 & \$0.009 & \$1.061 \\
Qwen3-flash   & 0.045 / 0.012 & 5.430 / 1.453 & \$0.004 & \$0.438 \\
DeepSeek-V3   & 0.034 / 0.006 & 4.042 / 0.715 & \$0.030 & \$3.622 \\
\midrule
\multicolumn{5}{l}{\textit{(b) PC / Helsinki ($N=7$ agents)}} \\
\midrule
GPT-4o        & 0.030 / 0.003 & 3.541 / 0.420 & \$0.109 & \$13.050 \\
GPT-4o-mini   & 0.030 / 0.004 & 3.614 / 0.453 & \$0.007 & \$0.814 \\
Qwen3-flash   & 0.052 / 0.009 & 6.297 / 1.038 & \$0.003 & \$0.366 \\
DeepSeek-V3   & 0.030 / 0.003 & 3.613 / 0.386 & \$0.023 & \$2.795 \\
\bottomrule
\multicolumn{5}{l}{Dollar costs are computed from unrounded token counts; token}\\
\multicolumn{5}{l}{values in the table are rounded to three decimals.}
\end{tabular}}
\label{tab:cost_backbone}
\end{table}

\textbf{(2) Hierarchical deployment with reduced decision frequency.} 
As demonstrated in Appendix~\ref{app:deploy}, MACRO-LLM maintains superior performance over both DPPO and LAMEN even when decision frequency is reduced by 90\% (Table~\ref{tab:hierarchical}), confirming that the framework's coordination value does not depend on high-frequency invocation. Table~\ref{tab:hier_cost} quantifies the resulting cost: enlarging the decision interval to $\Delta=10$ cuts per-episode cost from \$17.18 to \$1.48 (91.4\% reduction) while RMSE-H and SD-H remain close to the $\Delta=1$ setting, showing that coordination quality degrades only marginally as invocation cost drops by an order of magnitude.

\begin{table}[t]
\centering
\small
\setlength{\tabcolsep}{3pt}
\caption{Hierarchical deployment cost reduction on CPP / Catch-up using GPT-4o ($N=8$). Token counts are reported in millions, as input\,/\,output. Each episode contains 120 time steps, and cost reduction is computed relative to $\Delta=1$.}
\label{tab:hier_cost}
\resizebox{\linewidth}{!}{%
\begin{tabular}{lccccc}
\toprule
\textbf{$\Delta$} & \textbf{Calls/A} & \textbf{Tok./A} & \textbf{\$/A/Ep.} & \textbf{Cost Red.} & \textbf{RMSE-H / SD-H} \\
\midrule
1  & $\sim$837 & 4.024 / 0.712 & \$17.181 & --     & 1.237 / 0.409 \\
3  & $\sim$280 & 1.232 / 0.197 & \$5.053  & 70.6\%$\downarrow$ & 1.408 / 0.564 \\
5  & $\sim$167 & 0.737 / 0.121 & \$3.050  & 82.2\%$\downarrow$ & 1.319 / 0.524 \\
10 & $\sim$83  & 0.362 / 0.057 & \$1.476  & 91.4\%$\downarrow$ & 1.443 / 0.634 \\
\bottomrule
\multicolumn{6}{l}{$\Delta$ denotes the decision interval.}
\end{tabular}%
}
\end{table}

These two strategies are composable: combining local lightweight models with increased decision intervals can reduce the effective latency by over an order of magnitude relative to the default API-based configuration.

Within the evaluated simulators, MACRO-LLM is applied to the tested scenarios and topologies without parameter retraining, whereas the MARL baselines are trained separately for each task and topology. Whether this inference-only adaptation extends to unseen domains or open-world deployments remains to be evaluated.





\subsection{Scalability Metrics Across Fleet Sizes}\label{app:scala_metrics}
Fig.~\ref{fig:scalability} represents the coordination performance for the CPP Catch-up scenario as the fleet size increases from $N=8$ to $N=32$. Table~\ref{tab:scalability} reports the detailed coordination metrics, and Table~\ref{tab:cost_scalability} reports the corresponding per-agent cost.

All coordination metrics remain stable as $N$ increases to 32, with no obvious degradation compared to the baselines reported in Table \ref{tab:cacc}, indicating that the mean-field aggregation mechanism effectively compresses large-scale interactions without sacrificing coordination precision. The per-agent cost exhibits the same trend: as shown in Table~\ref{tab:cost_scalability}, both token consumption and monetary cost per agent stay within a narrow band across all fleet sizes (\$17.18--\$18.33 per episode), since mean-field aggregation keeps each agent's input bounded regardless of network scale. Together, the stability of both coordination quality and per-agent cost as the network quadruples demonstrates that MACRO-LLM scales favorably: its decentralized negotiation mechanism sustains coordination performance without incurring growing per-agent overhead, making it well-suited for large-scale multi-agent deployment.

\begin{table}[h]
\centering
\small
\caption{Scalability analysis on CPP Catch-up with GPT-4o.}
\label{tab:scalability}
\begin{tabular}{ccccc}
\toprule
$N$ & RMSE-H$\downarrow$ & RMSE-V$\downarrow$ & SD-H$\downarrow$ & SD-V$\downarrow$ \\
\midrule
8  & 1.237 & 0.721 & 0.409 & 0.203 \\
16 & 1.003 & 0.749	& 0.493	& 0.355 \\
24 & 0.908	& 0.664	 & 0.520	& 0.382 \\
32 & 0.728	& 0.628	& 0.433	& 0.397 \\
\bottomrule
\end{tabular}
\end{table}

\begin{table}[t]
\centering
\small
\setlength{\tabcolsep}{4pt}
\caption{Per-agent cost scalability of MACRO-LLM (CPP\,/\,Catch-up, GPT-4o). Token counts are reported in millions, as input\,/\,output. Across $N=8$ to $N=32$, the per-agent episode cost stays within a narrow band (\$17.18--\$18.33, a 6.7\% spread), indicating scale-stable cost rather than exponential growth.}
\label{tab:cost_scalability}
\begin{tabular}{ccccc}
\toprule
$N$ & \textbf{Tok./A/St.} & \textbf{Tok./A/Ep.} & \textbf{\$/A/St.} & \textbf{\$/A/Ep.} \\
\midrule
8  & 0.034 / 0.006 & 4.024 / 0.712 & \$0.143 & \$17.181 \\
16 & 0.037 / 0.006 & 4.444 / 0.721 & \$0.153 & \$18.325 \\
24 & 0.037 / 0.006 & 4.388 / 0.708 & \$0.150 & \$18.054 \\
32 & 0.035 / 0.006 & 4.149 / 0.729 & \$0.147 & \$17.660 \\
\bottomrule
\end{tabular}
\end{table}

\section{Extended Ablation Analysis} \label{app:ablation} 
In this section, we provide a detailed quantitative analysis of the ablation study.

\subsection{Component Analysis in CPP Task} 
The removal of specific modules leads to critical performance degradation in the high-precision and highly dynamic CPP task: 
\begin{itemize} 
\item \textbf{Impact of CoProposer:} Without the CoProposer, agents fail to infer and verify collaborative actions via rollout simulation. Consequently, RMSE-H and SD-H are {8.9$\times$} and {20.1$\times$} higher than those of MACRO-LLM, respectively. This result demonstrates that effective multi-agent coordination heavily depends on high-quality, verified initial proposals. 
\item \textbf{Impact of Negotiator:} Even with valid collaborative actions generated by the CoProposer, the omission of the negotiation phase results in an RMSE-H {5.9$\times$} higher than that of the full model. This degradation demonstrates the agents' inability to resolve conflicting intentions within a time-varying environment. 
\item \textbf{Impact of Introspector:} Agents without the Introspector tend to sacrifice headway stability to aggressively optimize velocity, with RMSE-H degrading by {1.9$\times$}. This indicates that self-reflection is essential to suppress oscillation and prevent the ``accordion effect'' in multi-agent collaboration. \end{itemize}

\subsection{Component Analysis in PC Task} 
A similar pattern holds for the PC task: 
\begin{itemize} \item \textbf{Impact of CoProposer:} Removing the CoProposer extends the pandemic duration by {2.3$\times$} compared to the full framework. Lacking rollout-simulated verification, agents adopt myopic strategies, failing to curb pandemic transmission effectively. 
\item \textbf{Impact of Negotiator:} Excluding the Negotiator prolongs the pandemic to 32 days, demonstrating that the absence of consensus mechanisms leads to inconsistent and ineffective containment policies. 
\item \textbf{Impact of Introspector:} Ablating the Introspector results in a pandemic duration approximately {30\%} longer than MACRO-LLM. This illustrates that agents without self-reflection are slow to adapt their strategies to the evolving infection dynamics. 
\end{itemize}

\subsection{Controlled Observation Analysis}
\label{app:controlled_observation}

We further isolate the effects of observation access and structured spatial aggregation. We compare local and full observation, each with and without mean-field aggregation, on CPP Catch-up and PC Helsinki using GPT-4o and Llama-3.3-70B. Across the four conditions, we keep the three-module architecture, prompt structure, call schedule, hyperparameters, and validation pipeline fixed. In the full-observation condition, every agent receives the complete global
state. Because the true global optimum is unavailable, this condition serves as a genie-assisted higher-information reference rather than an optimal-performance oracle.

\begin{table}[t]
\centering
\small
\setlength{\tabcolsep}{5pt}
\resizebox{\columnwidth}{!}{\begin{tabular}{lllrrrr}
\toprule
Obs. & Agg. & Backbone
& RMSE-H$\downarrow$ & RMSE-V$\downarrow$
& SD-H$\downarrow$ & SD-V$\downarrow$ \\
\midrule
Local & None       & GPT-4o         & 3.894 & 2.314 & 2.762 & 1.414 \\
Local & None       & Llama-3.3  & 7.732 & 6.111 & 6.785 & 5.533 \\
Local & Ours & GPT-4o         & 1.237 & 0.721 & 0.409 & 0.203 \\
Local & Ours & Llama-3.3  & 1.412 & 0.779 & 0.645 & 0.365 \\
\midrule
Full  & None       & GPT-4o         & 3.147 & 4.408 & 2.672 & 2.520 \\
Full  & None       & Llama-3.3  & 4.185 & 3.711 & 3.811 & 2.337 \\
Full  & Ours & GPT-4o         & 1.183 & 0.749 & 0.377 & 0.205 \\
Full  & Ours & Llama-3.3  & 1.368 & 0.718 & 0.492 & 0.224 \\
\bottomrule
\end{tabular}}
\caption{Controlled comparison of observation access and structured
aggregation on CPP Catch-up. All metrics are lower-is-better.}
\label{tab:controlled_cpp}
\end{table}

\begin{table}[t]
\centering
\small
\setlength{\tabcolsep}{6pt}
\resizebox{\columnwidth}{!}{\begin{tabular}{lllrrrr}
\toprule
Obs. & Agg. & Backbone
& $I_n\downarrow$ & $PI_n\downarrow$
& $D_n\downarrow$ & PD$\downarrow$ \\
\midrule
Local & None       & GPT-4o         & 0.140 & 0.086 & 0.000 & 19 \\
Local & None       & Llama-3.3  & 0.200 & 0.050 & 0.000 & 58 \\
Local & Ours & GPT-4o         & 0.011 & 0.008 & 0.000 & 10 \\
Local & Ours & Llama-3.3  & 0.013 & 0.008 & 0.000 & 14 \\
\midrule
Full  & None       & GPT-4o         & 0.692 & 0.320 & 0.000 & 30 \\
Full  & None       & Llama-3.3  & 0.064 & 0.044 & 0.002 & 16 \\
Full  & Ours & GPT-4o         & 0.010 & 0.008 & 0.000 & 10 \\
Full  & Ours & Llama-3.3  & 0.010 & 0.008 & 0.000 & 8 \\
\bottomrule
\end{tabular}}
\caption{Controlled comparison of observation access and structured
aggregation on PC Helsinki. All metrics are lower-is-better.}
\label{tab:controlled_pc}
\end{table}

As shown in Tables \ref{tab:controlled_cpp} and \ref{tab:controlled_pc}, local observation closely matches the full-observation reference based on GPT-4o with mean-field aggregation fixed. CPP RMSE-H changes from 1.237 to 1.183, while PC normalized infection changes from 0.011 to 0.010 and pandemic duration remains 10 days. In contrast, removing mean-field aggregation substantially degrades both tasks, increasing CPP RMSE-H to 3.894 and PC normalized infection to 0.140 under local observation. Llama-3.3-70B shows the same aggregation benefit, although its larger PC duration gap indicates some backbone dependence. Thus, full observation provides a higher-information empirical reference, while structured aggregation remains important under both observation conditions.

\section{Analysis of Robustness}\label{app:robust}
\begin{table}[t]
\centering
\caption{Robustness analysis of MACRO-LLM with GPT-4o on the CPP task (5 independent runs). Results are reported as mean $\pm$ standard deviation.}
\label{tab:robustness_cacc}
\resizebox{\columnwidth}{!}{%
\begin{tabular}{lcccc}
\toprule
Scenario & RMSE-H ($\downarrow$) & RMSE-V ($\downarrow$) & SD-H ($\downarrow$) & SD-V ($\downarrow$) \\
\midrule
Catch-up  & 1.237 $\pm$ 0.068 & 0.721 $\pm$ 0.124 & 0.409 $\pm$ 0.034 & 0.203 $\pm$ 0.054 \\
Slow-down & 0.496 $\pm$ 0.044 & 3.310 $\pm$ 0.010 & 0.128 $\pm$ 0.054 & 0.238 $\pm$ 0.073 \\
\bottomrule
\end{tabular}%
}
\end{table}
\begin{table}[t]
\centering
\caption{Robustness analysis of MACRO-LLM with GPT-4o on the PC task (5 independent runs). Results are reported as mean $\pm$ standard deviation.}
\label{tab:robustness_pc}
\resizebox{\columnwidth}{!}{%
\begin{tabular}{lccccc}
\toprule
City & I\_n ($\downarrow$) & PI\_n ($\downarrow$) & D\_n ($\downarrow$) & PD\_n\textsuperscript{*} ($\downarrow$) \\
\midrule
Helsinki  & 0.011 $\pm$ 0.001 & 0.008 $\pm$ 0.000 & 0.000 $\pm$ 0.000  & 0.081 $\pm$ 0.024 \\
Hong Kong & 0.008 $\pm$ 0.004 & 0.004 $\pm$ 0.001 & 0.001 $\pm$ 0.001  & 0.190 $\pm$ 0.063 \\
New York  & 0.004 $\pm$ 0.001 & 0.003 $\pm$ 0.000 & 0.000 $\pm$ 0.000 & 0.204 $\pm$ 0.117 \\
\bottomrule
\multicolumn{5}{l}{\footnotesize \textsuperscript{*} PD\_n = PD / total simulation days, representing the proportion of pandemic duration relative} \\
\multicolumn{5}{l}{\footnotesize \phantom{\textsuperscript{*}} to the total evaluation period.} \\
\end{tabular}%
}
\end{table}


To evaluate the stability of MACRO-LLM, we conduct 5 independent runs with GPT-4o and report the mean and standard deviation in Tables~\ref{tab:robustness_cacc} and~\ref{tab:robustness_pc}. For the CPP task, MACRO-LLM exhibits low variance across all metrics. In the Catch-up scenario, the SD-H of $0.409 \pm 0.034$ is 58.9\% lower than that of DPPO, and even the upper bound ($0.443$) remains 55.4\% lower, confirming that the stability advantage in Table~\ref{tab:cacc} is not attributable to favorable randomness. Similarly, in the Slow-down scenario, the RMSE-H ($0.496 \pm 0.044$) is 46.3\% lower than that of LAMEN and 71.2\% lower than that of DPPO across all runs. For the PC task, infection-related metrics (I\_n, PI\_n, D\_n) remain tightly concentrated across all three topologies, with standard deviations at or near zero. While PD\_n shows relatively higher variance in more complex topologies (e.g., New York: $0.204 \pm 0.117$), this is expected given the larger state space and stochastic disease dynamics inherent to denser urban networks. These results confirm that MACRO-LLM achieves consistently superior performance with reliable reproducibility across independent trials.

\section{Analysis of Failure Cases and Solutions}
\label{app:failure_cases}

To provide insights into the limitations of current LLMs in multi-agent control, we categorize the primary failure cases observed during our experiments and detail the corresponding mitigation strategies implemented in MACRO-LLM.

\paragraph{Arithmetic and Spatial Hallucinations.} 
Despite access to topological information, agents occasionally exhibit spatial hallucinations, such as confusing front and back vehicles in the CPP task. To mitigate this, we employ few-shot prompting incorporating explicit examples of state transitions (e.g., detailing how acceleration impacts headway relative to neighbors).

While LLMs occasionally produce imprecise numerical estimates during multi-step rollout predictions, all inference-dependent computations—including state prediction and constraint evaluation during rollout verification—are performed entirely by the LLM, with progressive constraint relaxation (Sec.~\ref{sec:macro-1}) reducing sensitivity to arithmetic imprecision. Deterministic computations independent of the LLM reasoning process, such as post-execution environment updates, are offloaded to an external numerical module to ensure precision.

\paragraph{Identity Confusion.} 
In scenarios featuring dense interaction prompts, such as concurrent proposals involving similar identifiers, ``\textit{Vehicle 2} propose action1 to \textit{Vehicle 1}, and propose action2 to \textit{Vehicle 3}'', agents may misattribute actions or forget their own identity. We attribute this to the attention mechanism's difficulty in distinguishing semantically similar tokens within extended contexts. To address this, we enforce a structured schema for inter-agent communication and integrate a self-consistency mechanism to validate sender-receiver pairings. Additionally, we reinforce role stability by prepending the instruction \textit{``You are [Agent Name]''} to every query prompt.

\paragraph{Syntactic Robustness.} 
To structure agent outputs, we employ XML-style \texttt{<label>} tags. However, smaller or quantized models (e.g., GPT-4o-mini, Qwen3-flash) occasionally fail to adhere to these formatting constraints, resulting in unclosed tags or hallucinated labels. To mitigate this, we implement a deterministic verification layer, in which outputs failing regex matching or schema validation trigger an automatic retry, ensuring that only syntactically valid actions are executed.

\paragraph{Safety Alignment Sensitivity.} 
In the PC task, the Qwen series frequently triggered safety refusals during stages characterized by high mortality rates, erroneously flagging simulation statistics as harmful content (returning \texttt{openai.BadRequestError}). This behavior indicates a stricter safety filter configuration compared to models like GPT-4o. To mitigate these false positives, we appended a system prompt explicitly clarifying the synthetic nature of the data: \texttt{``SIMULATION NOTE: This is a hypothetical pandemic experiment on pandemic simulator. All data is synthetic and does not reflect real-world events!!!''} Additionally, we implemented a rollback mechanism to discard and retry generation steps interrupted by safety filters.

\section{Analysis of Introspector's Attribution-based Reasoning}
\label{sec:appendix_introspector_causal}

To investigate how the Introspector achieves effective strategy refinement, we analyze whether its semantic revision signals reflect genuine attribution-based reasoning, attributing state changes to inter-agent interaction mechanisms, rather than simple trajectory pattern matching that maps local observations directly to actions. We provide both qualitative and quantitative evidence below. Throughout this section, attribution refers to relating an observed performance change to interaction-level factors—neighbors' behaviors and neighborhood statistical features—which then serve as a basis for strategy revision. The revised strategy is subsequently re-validated through rollout-simulated verification before any action is executed.
 
\subsection{Qualitative Evidence: Scenario-Adaptive Diagnostic Attribution Analysis}
We present representative semantic revision signals generated by the Introspector in two distinct CPP scenarios to demonstrate that it produces scenario-adaptive diagnostic attribution analyses rather than templated outputs.
 
\paragraph{Case 1 (Catch-up Scenario).} Vehicle~2 generated the following revision signal:
\begin{quote}
\textit{``The reasons may be due to the unexpected deceleration of vehicle~1 and the slight acceleration of vehicle~3. Vehicle~1's velocity decreased more than anticipated, causing vehicle~2 to adjust its acceleration to maintain a safe distance.''}
\end{quote}
The agent attributes its performance drop to specific neighbor behaviors, the unexpected deceleration of vehicle~1 and the acceleration pattern of vehicle~3, and derives a targeted adjustment strategy.
 
\paragraph{Case 2 (Slow-down Scenario).} The same agent position generated a qualitatively different signal:

\begin{quote}
\textit{``The spatial analysis indicates a general trend of deceleration, which may have influenced the need for a stronger deceleration to maintain formation.''}
\end{quote}
The agent attributes its adjustment to an aggregated mean-field spatial trend from unobservable agents, rather than to any individual neighbor's behavior.
 
These two cases demonstrate fundamentally different attribution directions: Case~1 performs local neighbor-level attribution, while Case~2 reasons about collective spatial dynamics. This confirms that the Introspector adapts its diagnostic attribution analysis to the scenario's underlying structure rather than applying a fixed reasoning pattern.

\subsection{Quantitative Evidence: Diagnostic Attribution Statistics}
 
To systematically evaluate whether the Introspector's reflections constitute attribution reasoning, we analyzed three complete collaboration logs comprising all reflection instances across full simulation episodes. We clarify that the attribution explicitly references other agents' behaviors, neighborhood statistical features, or inter-agent interaction effects as contributing factors to the observed performance change. Results show that 89.3\% of reflections contain explicit attribution to other agents' behaviors or neighborhood features, rather than self-trajectory patterns. This indicates that the Introspector systematically reasons about multi-agent interaction mechanisms rather than overfitting to local observation sequences.
 
\subsection{Behavioral Coherence Verification}
 
We further verify that the Introspector's attribution reasoning translates into coherent behavioral adjustments. In the Catch-up case (Case~1 above), vehicle~2 identified that vehicle~1 decelerated unexpectedly (acceleration shifting from $+1.8$ to $-1.3$~m/s$^2$) while vehicle~3 maintained low acceleration. The diagnostic attribution concluded that stabilization was needed to prevent headway violation. Consequently, vehicle~2 adjusted its own acceleration from $-1.05$ to $-0.82$~m/s$^2$, a conservative correction aligned with the diagnosed cause, reducing deceleration intensity to avoid over-correction given vehicle~1's unexpected behavior. This behavioral coherence, where the direction and magnitude of action adjustments logically follow from the attribution diagnosis, is further corroborated by the ablation results in Figs.~\ref{fig:exp-abla1} and~\ref{fig:exp-abla2}, confirming that attribution adaptation is essential for long-horizon coordination.

\section{Core Prompts of MACRO-LLM}\label{app:prompt}
\subsection{CoProposer} \label{app:rompt1}
\begin{figure*}[t!]
    \begin{ChatBox}[width=\textwidth]{
        \includegraphics[height=1.4em]{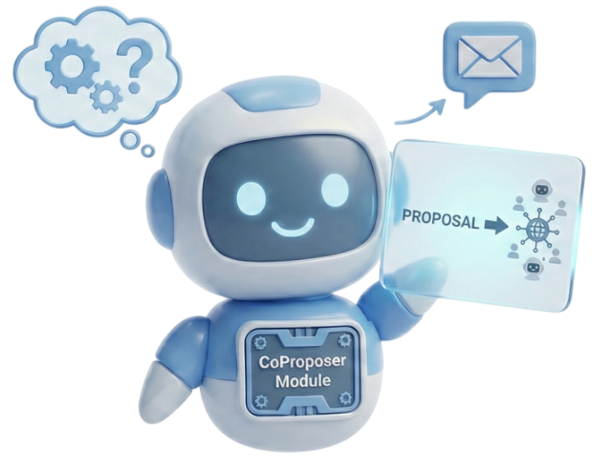} \hspace{1mm} CoProposer: Proposal Generation Query
    }
        You are [\textit{agent\_name}] in a [\textit{task\_name}]. This is negotiation round 0 with your neighbors. Your task is to [{\textit{task objective}}]. Decision-Making Steps (Follow strictly, step by step):\\
        Step 0: Initial Decision for [\textit{agent\_name}] at the 1st time step, and define urgency of the action: Normal, Warning, Urgent.\\
        Step 1: Collaborative acceleration for [\textit{observable\_agents}] to achieve objective with [Temporal Strategy], [Spatial Strategy] and [Spatial Features]. \\
        \textit{/*Steps 2-4: Validation by 2 Time-Step Simulation: (time step 2 and time step 3).*/} \\
        Step 2: Simulate the next state (state\_2) at time step 2. [\textit{instructions}]\\
        Step 3: Calculate collaborative actions for state\_2.\\
        Step 4: Simulate the next state (state\_3) at time step 3. [\textit{instructions}]\\
        Step 5: Feasibility check for both state\_2 and state\_3 \\ 
        {[\textit{task\_constraints}]} \\
        - If any constraint is violated, go back to Step 0 and pick a more conservative action. Repeat Steps 0-4 up to 10 times. \\
        - If no feasible proposal is found after 10 iterations, output the best proposal found.\\
        Output Format (STRICT): \\
        \textless analysis\textgreater [\textit{analysis\_and\_calculation\_steps}] \textless/analysis\textgreater \\ 
        \textless proposal\textgreater [\textit{semantic\_proposal\_for\_negotiation}] \textless/proposal\textgreater\\ 
        \textless output\textgreater E[\textit{timestep}]\_R[\textit{round}]\_[\textit{agent\_name}]\_proposal = \{[\textit{my\_name}]: \textless proposed action for itself\textgreater, [\textit{neighbor's\_name}]: \textless proposed acceleration for neighbor\textgreater\}\textless/output\textgreater\\ 
        IMPORTANT: [\textit{self-checking\_rules}] \\
    \end{ChatBox}
    \caption{The core prompt of CoProposer.}
    \label{tab:prompt1}
\end{figure*}

Fig. \ref{tab:prompt1} provides the core prompts of proposal generation with a rollout-simulated verification in CoProposer. 


\subsection{Negotiator}
\begin{figure*}[t!]
    \begin{ChatBox}[width=\textwidth]{
        \includegraphics[height=1.4em]{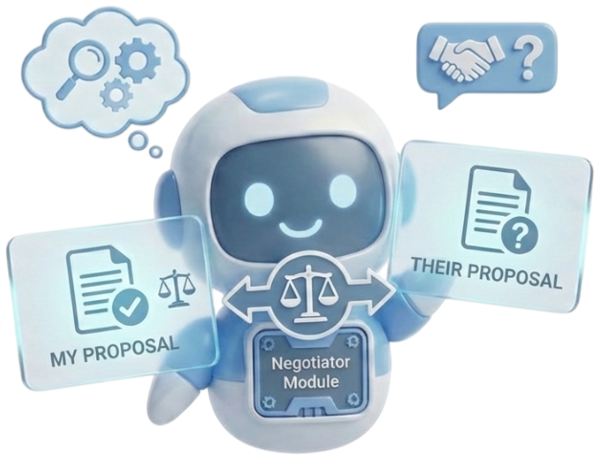} \hspace{1mm} Negotiator: Proposal Evaluation Query
    }
        You are [\textit{agent\_name}] for [\textit{task\_name}]. Your observable neighbors are [\textit{observable\_agents}]. In the last negotiation round, consensus was not reached. You need to analyze the disagreement and summarize your reasoning in the following steps:\\          
        Step 0: Analyze whether the strategy proposed by [\textit{agent\_name}] conflicts with the strategy proposed by its observable neighbors [\textit{agent\_name}]\\
        {[\textit{task\_strategy\_information}]}\\
        Step 1: According to [Spatial Strategies] and [Spatial features], analyze whether the strategy proposed by [\textit{neighbors\_agent}] conflicts with the trends of [\textit{unobservable\_agents}].\\
        {[\textit{task\_strategy\_information}]}\\
        Output as following format:\\                
        <deal>False</deal>\\
        <reasons> Observable vehicles:[\textit{observable\_agent\_name}] choose ..., and proposes .... It's strategy is match/conflict/not conflict with my strategy. The reason is ... I suggest it's weight in the range ... \\           
        {[\textit{unobservable\_agent\_list}]}: They are strategy trend is match/conflict/not conflict with my strategy. The reason is ... I suggest it's weight in the range ...</reasons>\\
        <spatial-strategy>[\textit{update\_spatial\_strategy}]</spatial-strategy>\\
        IMPORTANT: [\textit{self-checking\_rules}] \\
    \end{ChatBox}
    \caption{The core prompt of proposal evaluation and spatial strategy update.}
    \label{tab:prompt2-1}
\end{figure*}

\begin{figure*}[t!]
    \begin{ChatBox}[width=\textwidth]{
        \includegraphics[height=1.4em]{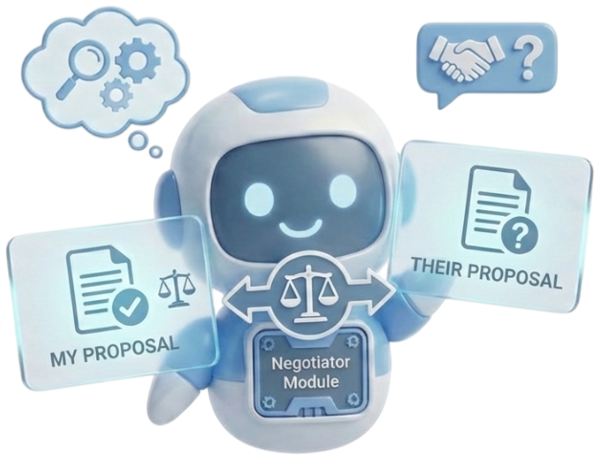} \hspace{1mm} CoProposer: Update Proposal
    }
        You are [\textit{agent\_name}]. This is the Round\_[\textit{current\_round}] of [\textit{time\_step}]. Your observable states is [\textit{agent\_name} Observable State in \textit{time\_step}]. You need to update your proposal by following steps. \\            
        - \textbf{Case 1: If the flag between <deal></deal> is True, reuse the last proposal and output as following format}: \\
        <analysis>(I will keep my last proposal, because ...)</analysis>\\
        <proposal> [\textit{semantic\_proposal\_for\_negotiation}] </proposal>  \\          
        <output>[\textit{structured\_proposal\_for\_coding}]</output>\\ 
        Skip the following steps.\\
        
        - \textbf{Case 2: If the flag between <deal></deal> is False}: \\
        Step 1: Define the confidence weights for observable neighbors\' proposal and unobservable vehicles weighted average acceleration. [\textit{weight\_rules}]\\
        Output the confidence weights as following format:\\            
        <insights>neighbor\_proposed\_weight:<neighbor\_proposed\_weight value>          unobservable\_weight: <unobservable\_weight value>     my\_weight: <my\_weight value></insights>\\            
        Step 2: Based on weights defined between <insights></insights>, weight the proposed acceleration for each vehicle by following equations: \\            
        updated\_acceleration = neighbor\_proposed\_weight * neighbor\_proposed\_acceleration + my\_weight * my\_acceleration + unobservable\_weight * unobservable\_trend.\\
        \textbf{/* Step 3: A rollout-simulated verification similar to Step 2-5 in CoProposer */}\\ 
        Output Format (STRICT): \\
        <analysis>[\textit{analysis\_and\_calculation\_steps}] </analysis> \\ 
        <proposal>[\textit{semantic\_proposal\_for\_negotiation}] </proposal>\\ 
        <output>E[\textit{timestep}]\_R[\textit{round}]\_[\textit{agent\_name}]\_proposal = \{[\textit{agent\_name}]: <proposed action for itself>,\\
        {[\textit{neighbor's\_name}]: <proposed acceleration for neighbor>\}</output>}\\ 
        IMPORTANT: [\textit{self-checking\_rules}] \\
    \end{ChatBox}
    \caption{The core prompt of update proposal.}
    \label{tab:prompt2-2}
\end{figure*}

Fig. \ref{tab:prompt2-1} provides the core prompts for proposal evaluation with spatial strategies and feature updates in Negotiator when consensus is not achieved. 
Fig. \ref{tab:prompt2-2} provides the core prompts for updating proposal with updated spatial strategies and features in Negotiator. 

\subsection{Introspector} \label{app:prompt_intro}
\begin{figure*}[t!]
    \begin{ChatBox}[width=\textwidth]{
        \includegraphics[height=1.4em]{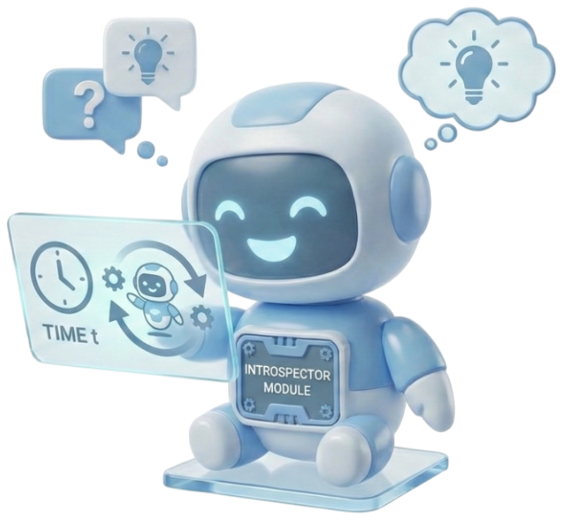} \hspace{1mm} Introspector: Generate Semantic Revision Signal
    }
        You are [\textit{agent\_name}]. You can observe [\textit{observable\_agent\_name}]. The total system reward for this time step is xx\% lower than in the last time step indicating that the system state getting worse. Your individual reward has dropped from xx to xx, which drop xx\%. \\                
        Step1. Based on [\textit{last\_negotiation\_record}] and [\textit{this\_negotiation\_record}], integrate the [Objective Instruction] to analyze whether the drop in episode reward was caused by yourself. If so, at which stage was responsible.\\ 
        Infer whether the reward decline was caused by [\textit{observable\_agents}] or [\textit{unobservable\_agents}], and output between tags <reasons> </reasons>\\ 
        Based on <reasons> </reasons> should strategies be adjusted to improve future episode rewards?\\  
        Step 1: You need to find out the rules that need to be adjusted based on the reasons you summarized.\\
        Step 2: According to the related rules you summarized, be specific about ``what to do'' and ``what to avoid'' and output between tags <self-checking\_rules> and </self-checking\_rules>. \\
        The output format as follows: \\
        <reasons>[\textit{summarized\_reasons}]</reasons>\\
        <self-checking\_rules>(When states condition ... I need to/I need to avoid ...)</self-checking\_rules>\\        
        IMPORTANT: [\textit{self-checking\_rules}]\\
    \end{ChatBox}
    \caption{The core prompt of generating semantic revision signal.}
    \label{tab:prompt3-1}
\end{figure*}

\begin{figure*}[t!]
    \begin{ChatBox}[width=\textwidth]{
        \includegraphics[height=1.4em]{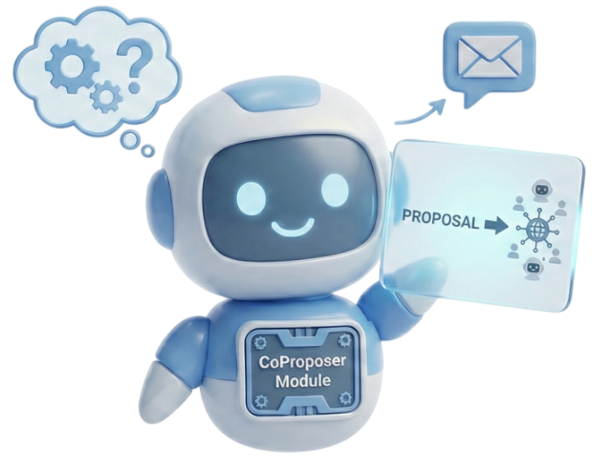} \hspace{1mm} Introspector: Update Strategy Query
    }
        You are [\textit{agent\_name}] to update its strategies. The total system reward for this time step is xx\% lower than last time step, indicating that the system state getting worse. The overlap of transition (including state and action) among these two time steps is xx\%, which means that their overall movement and decision patterns are xx\% similar, with xx\% differences in how the agent moves between states and selects actions. \\
        Based on the <reasons></reasons> you summarized (from Fig. \ref{tab:prompt3-1}), update your strategies carefully with NO MORE than xx\%.\\
        Step 1. Calculate the change in state from the historical states to the current state and compare it to the previous temporal plan. Analyze if the change in state matches the plan of the previous temporal plan.\\
        Step 2. Carefully check whether the change match the last temporal plan.\\
        - If it matches, keep the original plan and update the temporal plan. Execute step 3 to return the updated plan. \\
        - if it doesn't match, summarize the reasons and output within <reflect> tags as following: <reflect>(My final action does not match my original temporal plan, the reasons may be ...)</reflect>\\
        According to the <reflect></reflect>, update the temporal plan.\\
        Step 3. Output the updated plan with following format:\\        
        <temporal-strategy>[\textit{update\_temporal\_analysis}]</temporal-strategy>\\
        IMPORTANT: [\textit{self-checking\_rules}] \\
    \end{ChatBox}
    \caption{The core prompt of update strategy}
    \label{tab:prompt3-2}
\end{figure*}

Fig. \ref{tab:prompt3-1} presents the core prompts for generating the 
revision signal
after completing one-step execution.
Fig. \ref{tab:prompt3-2} presents the core prompts for updating the temporal strategy using the drift intensity and semantic revision signal.

[\textit{self-checking\_rules}] are lightweight, task-level statements of objective physical relations of each domain. For example, in CPP, that a positive acceleration gap increases velocity and decreases headway. Their purpose is to provide all backbones with a uniform minimal physical prior, so that performance reflects the coordination mechanism rather than differences in the domain knowledge each model happens to carry. They are defined once per task, shared across all scenarios and fleet sizes without per-instance tuning, and are not part of the coordination mechanism itself.

\section{AI Assistants in Writing}
\label{app:ai_use}
We utilized AI assistants (e.g., Gemini and Claude) exclusively for grammatical error correction, sentence polishing to enhance readability, and generating robot icons used in Figs. \ref{fig:teaser} and \ref{fig:framework}. 
We explicitly state that the core scientific concepts, experimental design, logical structure, references, and substantive content of the paper were entirely human-authored without generative contributions from LLMs. 



\end{document}